%% file: main.tex
\documentclass[preprint,5p,twocolumn]{elsarticle}

\usepackage[utf8]{inputenc}
\usepackage[english]{babel}
\usepackage{graphicx, subfig}
 
\usepackage{caption}
\usepackage{float}
\usepackage{color}

\usepackage{amsmath}
\usepackage{hyperref}

\usepackage{flushend}

\usepackage{algorithm}
\usepackage{algorithmic}

\usepackage{placeins}
\usepackage{booktabs}

\usepackage{nth}
\usepackage{acronym}
\usepackage[capitalise]{cleveref}

\usepackage{lipsum}
\newcommand\blfootnote[1]{%
  \begingroup
  \renewcommand\thefootnote{}\footnote{#1}%
  \addtocounter{footnote}{-1}%
  \endgroup
}

\graphicspath{ {fig/} }

\newacro{EHV}{Extra-High-Voltage}
\newacro{HV}{High-Voltage}
\newacro{MV}{Medium-Voltage}
\newacro{LV}{Low-Voltage}
\newacro{DER}{Distributed Energy Resources}
\newacro{RES}{Renewable Energy Sources}
\newacro{OPF}{Optimal Power Flow}
\newacro{DSO}{Distribution System Operator}
\newacro{TSO}{Transmission System Operator}
\newacro{HPC}{High Performance Computing}
\newacro{PV}{Photovoltaic}
\newacro{DAG}{directed acyclic graph}
\newacro{ANN}{Artificial Neural Network}

\newacro{OLTC}{On-Load Tap Changer}
\newacro{HPC}{High-Performance Computing}
\newacro{MCS}{Monte Carlo Simulation}
\newacro{HT}{Hyper-Threading}

\newacro{ALU}{arithmetic logic unit}
\newacro{JIT}{just-in time}

\newacro{AC}{Alternating Current}
\newacro{PF}{Power Flow}
\newacro{NR}{Newton-Raphson}
\newacro{PPF}{Probabilistic Power Flow}

\newacro{FS}{Forward Substitution}
\newacro{BS}{Backward Substitution}
\newacro{SMT}{Simultaneous multithreading}
\newacro{SIMD}{Single Instruction Multiple Data}
\newacro{SIMT}{Single Instruction Multiple Threads}

\newacro{CRS}{Compressed Row Storage}
\newacro{CCS}{Compressed Column Storage}
\newacro{FP}{floating-point arithmetic}

\newacro{JM}{update jacobian matrix}
\newacro{NPM}{Nodal Power Mismatch}
\newacro{SM}{Stream Multiprocessor}
\newacro{SP}{Stream Processor}

\newacro{VMAD}{vector multiply and add}
\newacro{MAD} {Multiply and Add}
\newacro{G-P}{Gilber-Peierls left-looking algorithm}

\journal{Sustainable Energy, Grids and Networks}

\begin{document}
	\begin{frontmatter}
		\title{Fast Parallel Newton-Raphson Power Flow Solver for Large Number of
		System Calculations with CPU and GPU}
		
		\author[e2n,iee]{Zhenqi Wang\corref{cor1}}
		\ead{zhenqi.wang@uni-kassel.de}
		\author[e2n,iee]{Sebastian Wende-von Berg}
		\author[e2n,iee]{Martin Braun}
		\cortext[cor1]{Corresponding author}
		
		\address[e2n]{Department of Energy Management and Power System Operation,\\ University of Kassel,\\ Wilhelmshöher Allee 73, Kassel, 34121, Germany}
		\address[iee]{Fraunhofer Institute for Energy Economics and Energy System Technology,\\ Königstor 59, 34119 Kassel, Germany}
				
		\begin{abstract}
			To analyze large sets of grid states, e.g. when evaluating the impact from the uncertainties of the renewable generation with probabilistic Monte Carlo simulation or in stationary time series simulation, large number of power flow calculations have to be performed. For the application in real-time grid operation, grid planning and in further cases when computational time is critical, a novel approach on simultaneous parallelization of many Newton-Raphson power flow calculations on CPU and with GPU-acceleration is proposed. The result shows a speed-up of over x100 comparing to the open-source tool pandapower, when performing repetitive power flows of system with admittance matrix of the same sparsity pattern on both CPU and GPU. The speed-up relies on the algorithm improvement and highly optimized parallelization strategy, which can reduce the repetitive work and saturate the high hardware computational capability of modern CPUs and GPUs well. This is achieved with the proposed batched sparse matrix operation and batched linear solver based on LU-refactorization. The batched linear solver shows a large performance improvement comparing to the state-of-the-art linear system solver KLU library and a better saturation of the GPU performance with small problem scale. Finally, the method of integrating the proposed solver into pandapower is presented, thus the parallel power flow solver with outstanding performance can be easily applied in challenging real-life grid operation and innovative researches e.g. data-driven machine learning studies.
		\end{abstract}
		
		
	\end{frontmatter}

	\section{Introduction}
	\blfootnote{©2021. This manuscript version is made available under the CC-BY-NC-ND 4.0 license http://creativecommons.org/licenses/by-nc-nd/4.0/}
	\input{intro.tex}

	\section{Review of parallelization on CPU and GPU}
	\label{sec:hardware_cpu_gpu}
	\input{hardware_cpu_gpu.tex}

	\section{Approach for a parallel NR-PF solver}
	\label{sec:implementation_cpu_gpu}
	\input{implementation_cpu_gpu.tex}

	\section{Batched Linear System Solver}
	\label{sec:batched_linear_solver}
	\input{batched_linear_solver.tex}

	\section{Case Studies and Performance Evaluation}
	\label{sec:case_study}

\input{case_study.tex}

	\section{Conclusion}
	\label{sec:conclusion}
	\input{conclusion.tex}

	\section*{Acknowledgements}
	The authors would like to thank Florian Schäfer and Dr. Alexander Scheidler for their suggestions to improve the quality of this paper. The work was supported by the European Union's Horizon 2020 research and innovation programme within the project EU-SysFlex under grant agreement No 773505.

	\bibliographystyle{elsarticle-num}
	\bibliography{paper}

\end{document}

%% file: intro.tex
The penetration of \ac{DER} e.g. Wind and PV causes high uncertainties in planning and operation of power systems. For both grid planning and operation, with the probability distribution of the infeed and load with uncertainties known, it can be formulated as \ac{PPF}-problem and evaluated with \ac{MCS}-\ac{PF}. The computational effort to solve the \ac{PPF} problem varies according to the complexity of the unknown variable and the sampling methods. In the past decades, the researches have succeeded in reducing the required number of \ac{PF}s \cite{Yu.2009, Su.2005, Usaola.2010}. Besides, the \ac{MCS}-\ac{PF} finds good application on a similar evaluation for the uncertainty with historical or synthetic injection and load profiles.

Static N-1 contingency analysis is required for the real-time and future grid state in the grid operation, with which grid congestion caused by the renewable energy needs to be properly handled with market and operational measurements \cite{Schafer.29.10.201831.10.2018}. Thus, it is essential to evaluate large number of \ac{PF}s fast. 

Furthermore, recent power system research trends show an increasing demand for a powerful \ac{PF} solver for many \ac{PF}s. Data-driven machine learning methods show great potential in power system state estimation, approximation of \ac{PF} and using \ac{ANN} to assist decision making in dispatch actions in grid operation. Those approaches require large number of \ac{PF}s of similar grid topology to be performed in the training phase \cite{Schafer.29.10.201831.10.2018, Mestav.2019, Liu.2019}. The proposed method finds successful application in the study of using \ac{ANN} for \ac{MV} power system state estimation \cite{Menke.2019}.

With the open source tools e.g. MATPOWER\cite{Zimmerman.2011} and pandapower\cite{Thurner.2018}, the grid can be easily modelled and single \ac{PF} can be conveniently solved. The performance of these tools in solving many \ac{PF}s with similar grid topology is unsatisfactory. Emerging works have shown great potential of the application of massive parallel hardware e.g. GPU in acceleration single \ac{PF}\cite{Araujo.2019, Guo.19.08.201222.08.2012, Su.2020, Li.2017} as well as many \ac{PF}s e.g. for static contingency analysis and online \ac{PPF} \cite{Roberge.2017, Zhou.2017b, Zhou.2018, Abdelaziz.2017, Abdelaziz.2018, Huang.2018b}.

Motivated by the aforementioned challenging use cases, a general-purpose parallel \ac{PF} solver for solving many \ac{PF}s for grid with same sparse pattern of admittance matrix is presented in this work. \ac{PF} with classical \ac{NR} method is efficiently solved on both CPU and GPU with the proposed approach. Recent development \cite{Zhou.2017b, Zhou.2017, Huang.2018} show the advantage of parallelization for solving many \ac{PF}s through batched operation on GPU. Our work distinguished itself by furthering to study the possible bottlenecks in the \ac{NR} algorithm. Through further optimization, the repetitive work can be reduced and the utilization rate of available computational resources can be maximized.

The main contributions of our work include the following. First, to optimize the performance on CPU, besides the task level parallelization on CPU, a new parallelization scheme on CPU with the extra explicit \ac{SIMD} parallelization \cite{Fog.052020}, which is the first of its kind to our best knowledge as revealed in our literature overview. The multi-threaded \ac{SIMD} LU refactorization shows further speed up comparing to the state-of-the-art KLU\cite{Davis.2010} library with task-level parallelization. Secondly, an easy-to-implement row-level parallelization strategy is proposed for batched sparse matrix operation on GPU. Thirdly, an improved GPU LU refactorization based on the latest advances \cite{Chen.2015,Zhou.2017} is proposed, which can increase the hardware saturation through the final stages in the LU-refactorization process and thus improve the performance on small batch size. Furthermore, the forward substitution backward substitution step is optimized with fine-grained parallelization strategy. Last but not least, the method of integrating the parallel \ac{PF} solver into the python-based open-source power system analysis tool is presented, which is essential of the application into real-life grid planning, operation and researches.

This paper is formulated in 5 sections. \cref{sec:hardware_cpu_gpu} introduces the CPU and GPU architecture and how the performance of computation tasks can be optimized respectively. \cref{sec:implementation_cpu_gpu} introduces the proposed approach of implementation of the parallel \ac{PF} solver. \cref{sec:batched_linear_solver} introduces the proposed batched linear solver for CPU and GPU platform. \cref{sec:case_study} presents the benchmarking result with comprehensive analysis. 


%% file: hardware_cpu_gpu.tex
\begin{figure}[t]
  \centering
  \includegraphics[width=0.48\textwidth]{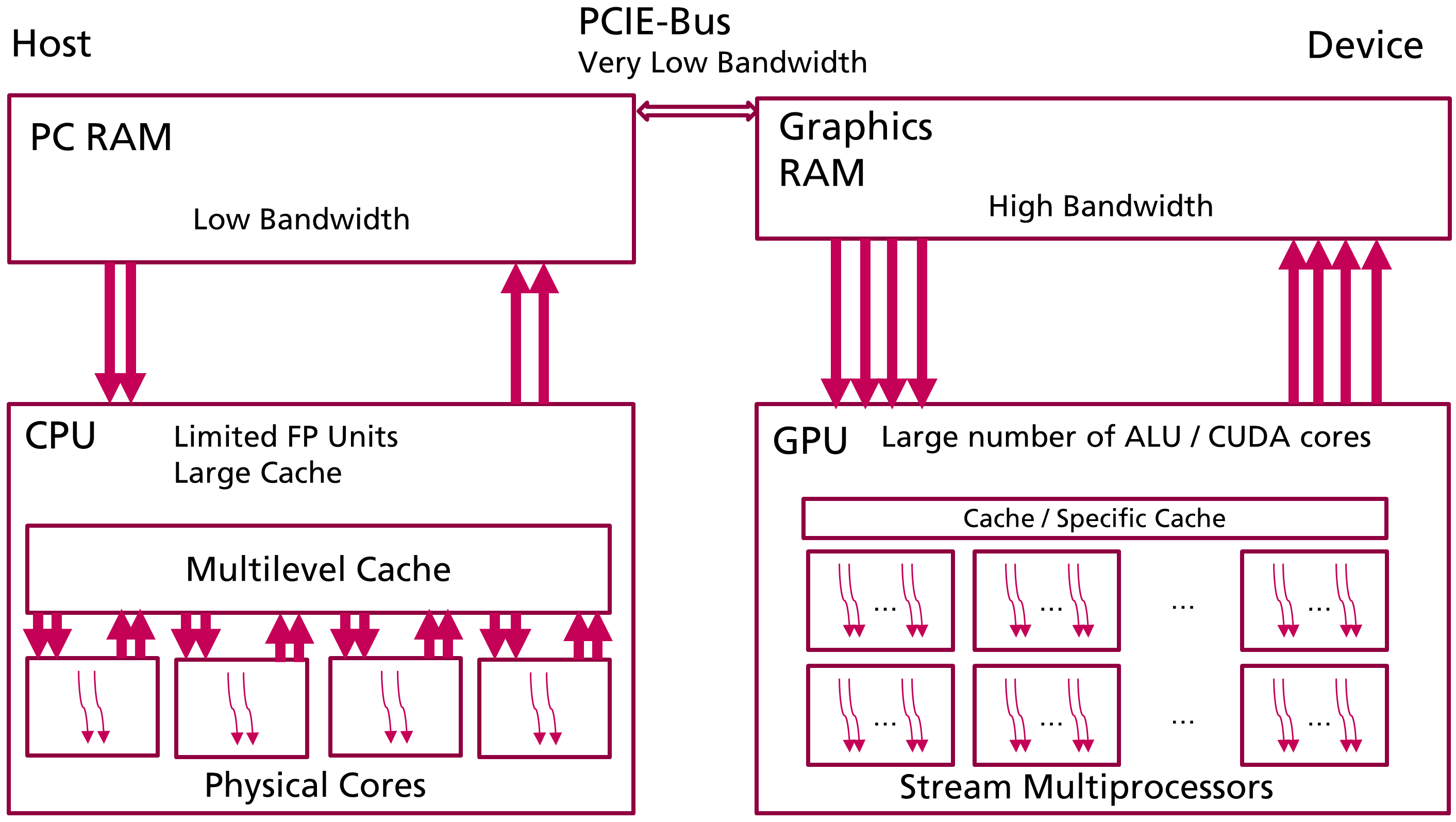}
  \caption{CPU-GPU architecture overview.}
  \label{fig:hardware_cpu_gpu}
\end{figure}

This section gives an overview of the modern CPU and GPU regarding its specialties in order to solve many \ac{PF}s efficiently. \cref{fig:hardware_cpu_gpu} shows a generalized CPU-GPU hardware structure. On modern CPU, multiple physical cores are usually available, with which multiple tasks can be performed independently. For computational tasks, a good utilization of multi-cores greatly improves the performance. Furthermore, with large size of on-chip cache available, the memory transaction with limited memory bandwidth can be avoided on cache-hits.

The \ac{SIMD} instruction set is available on CPU on each physical core, which can execute the same operation on multiple data resides in the extended registers with the size of 128 bits of higher \cite{Gepner.2017}. This corresponds to 2 and more double precision \ac{FP}s, with which the overall \ac{FP} capability of CPU is greatly improved \cite{EvguenyKhartchenko.012018}. The SIMD instruction set e.g. SSE and AVX2 is available on most modern CPU, which can carry out 2 and 4 double-precision \ac{FP}s simultaneously.

Comparing to CPU, modern GPU is designed not only for graphic rendering but also as a powerful highly parallel programmable processor \cite{Owens.2008}. Large number of parallel \ac{SP}s are integrated on-chip, which result in a high \ac{FP}s peak performance comparing to CPU. The \ac{SP}s are clustered into \ac{SM}s, which contains L1-cache, shared memory and control units \cite{NVIDIA.}. Besides, GPU has high memory bandwidth as shown in \cref{fig:hardware_cpu_gpu}, which brings advantage on the memory-bounding tasks.

\subsection{Parallelization on CPU}

Multi-threaded task-level parallelization can be realized with OpenMP \cite{Dagum.1998}, which is a popular library in scientific computing for parallelization on shared-memory. This parallelism paradigm is called \ac{SMT}\cite{D.M.Tullsen.1995}. OpenMP uses compiler directives to allocate the parallelized regions in the serialized program during the compiling phase.

Since OpenMP 4.0, \ac{SIMD} parallelization is directly supported within the framework \cite{OpenMPArchitectureReviewBoard.072013}. On top of the \ac{SMT} parallelization, mini-batch tasks can further profit from the usage of SIMD instruction set, this type of parallelism is denoted as SMT+SIMD.

\subsection{Principle of SIMT parallelization on GPU}

\begin{figure}[t]
  \centering
  \includegraphics[width=0.42\textwidth]{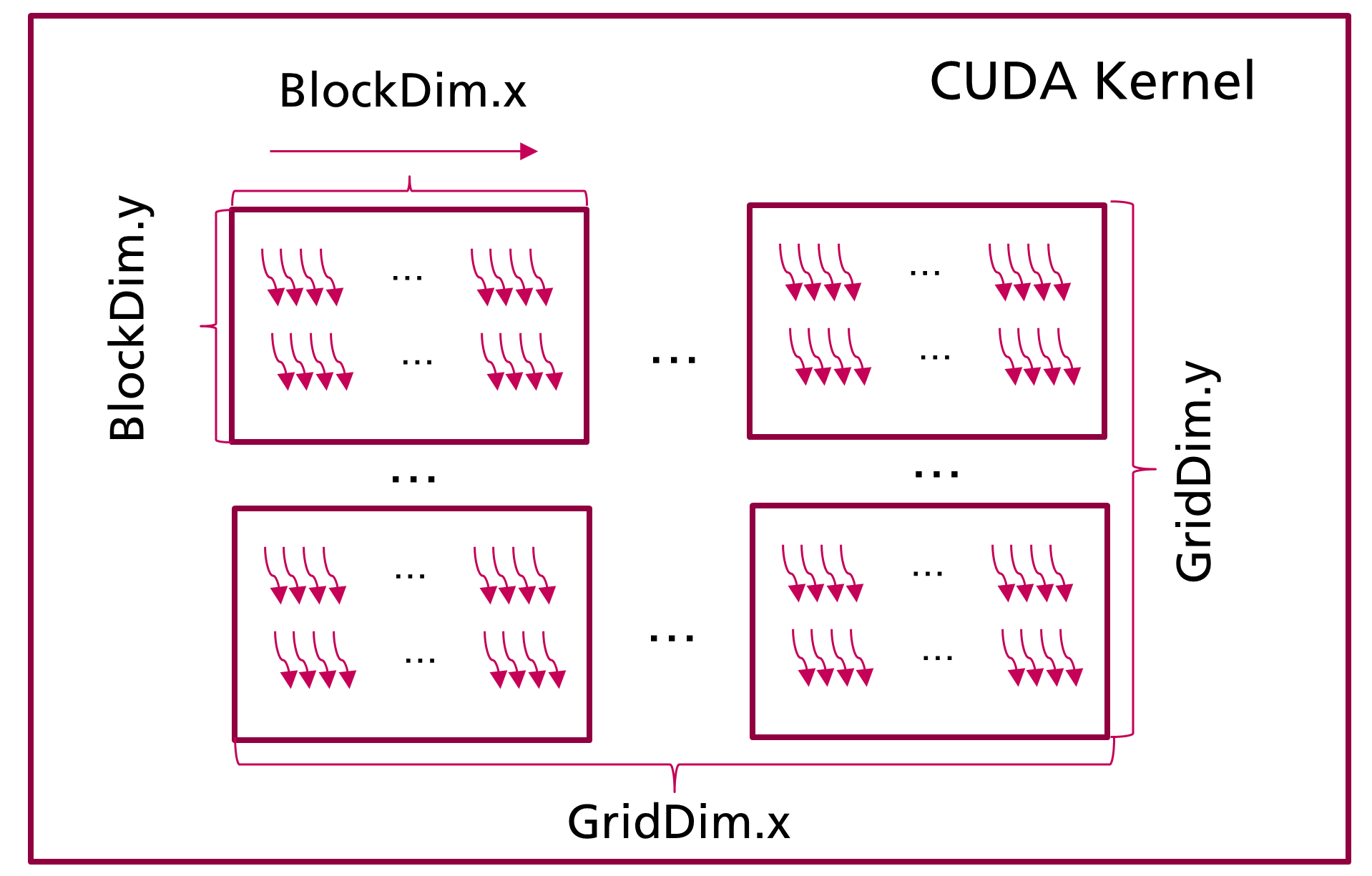}
  \caption{Simplified model of CUDA kernel structure.}
  \label{fig:cuda_kernel_structure}
\end{figure}

In order to perform large-scaling scientific computation tasks on massive parallelization hardware like GPU, \ac{SIMT} programming model is widely adapted. Using NVIDIA CUDA as an example, the user is able to program functions called CUDA Kernel, which defines the behavior of massive simultaneous threads. A simplified 2D programming model of CUDA kernel is shown in \cref{fig:cuda_kernel_structure}. The massive threads are organized in a two-level structure, with each level supports up to 3-dimensions. The first level is called thread block, all the threads within a block are executed on a single \ac{SM} and can be synchronized explicitly with synchronization barrier \cite{NVIDIA.}. All the thread blocks are organized in a grid. The execution of grid is distributed on all the \ac{SM}s on-chip. The thread is aware of its indexing in the two-level structure, which is essential to define the data-indexing relevant individual tasks. The size of thread block and grid can be configured upon kernel call.

During the execution of a CUDA kernel, 32 consecutive threads within one block are aggregated into the minimum execution unit called warp \cite{Zhang.12.02.201116.02.2011}. The threads within one warp execute the same instruction. When threads within a warp encounter different instructions, these are executed in a serial mode, thus lose the advantage of parallelism. To fully utilize the calculation capability for computing-bounding tasks, the first criterion for GPU performance optimization is to avoid thread divergence within one warp. 

Optimizing memory operation with the on-device memory is another important factor. Due to the high-latency of the graphics memory (multiple hundreds of cycles) \cite{Volkov.2016}, the instruction of memory transaction is queued in a First-In-First-Out way. It is possible to hide this latency and saturate the memory bandwidth by keeping the queue busy \cite{Zhang.12.02.201116.02.2011}. For each memory transaction, 32-bytes data of consecutive memory sections is  accessed, to maximize the efficiency of memory transaction and reduce excessive memory transaction, it is important to enable the warp access for coalesced memory, especially for memory-bounding tasks \cite{NVIDIA.November2019}. Constants and texture can be accessed through specific cache, which improves the access with sparse matrix indexing. 

Any tasks on GPU can only be executed, when the data is available on the device memory. \cref{fig:hardware_cpu_gpu} shows the limited bandwidth between CPU (Host, H) and GPU (Device, D). To optimize this, the required data transaction between host and device should be possibly reduced. Furthermore, CUDA offered the possibility of overlapping different operations (H2D, D2H memory transaction and kernel execution) through the usage of CUDA stream \cite{Huang.2018b}.

%% file: implementation_cpu_gpu.tex
\begin{figure*}[t]
  \centering
  \includegraphics[width=0.95\linewidth]{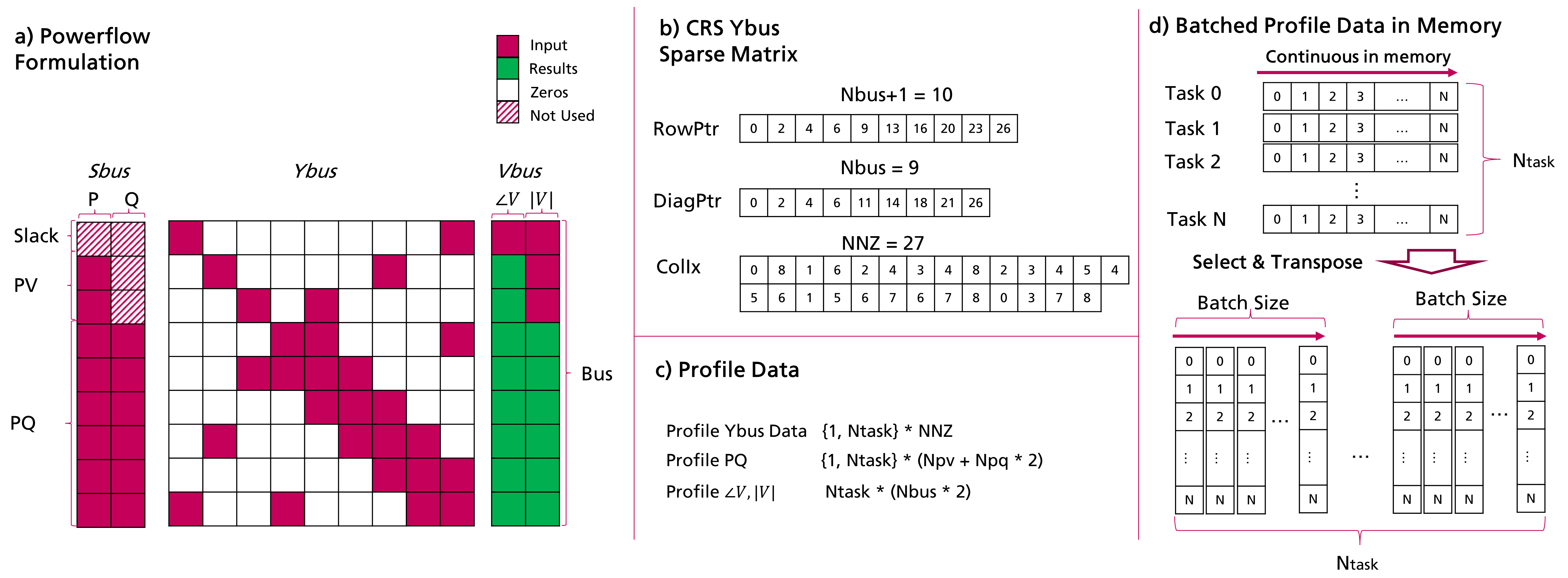}
  \caption{CRS sparse matrix for \ac{NR}-\ac{PF} and batched data.}
  \label{fig:form_nr}
\end{figure*}

In this work, an approach of the acceleration of many \ac{NR}-\ac{PF}s with same sparsity pattern admittance matrix $Y_{bus}$ (according to the MATPOWER convention \cite{Zimmerman.2011}) is presented. This special property leads to the same sparsity pattern of the \ac{JM} required in the iterative solving process, which brings further advantages for speed-up by reducing repetitive works as following:
\begin{itemize}
  \item Reuse of indexing of sparse matrix
  \item Reuse of static lookup for sparse matrix update 
  \item Reuse of memory working space
\end{itemize}

\cref{fig:form_nr} shows exemplary the indexing data and profile data required for the parallel \ac{PF} solver. The $Y_{bus}$ is stored in \ac{CRS}-format with extended diagonal index. The standard \ac{CRS} sparse matrix indexing consists of $RowPtr$ with the length of number of rows plus one and $ColIx$ equals the number of non-zero elements, which is efficient for the iteration over rows of the matrix. The extra $DiagPtr$ to represent the data index of the diagonal element of each row for the convenience of iterating and aggregating on the diagonal elements (required in update \ac{JM}). \cref{fig:form_nr}b gives the aforementioned sparse indexing of the non-zero elements of $Y_{bus}$ in \cref{fig:form_nr}a. For different types of calculation, the variable profile data required is shown in \cref{fig:form_nr}c. For PPF, $Y_{bus}$ requires only once while the $PQ$ equals the size of the sampling. For static N-1 analysis, the number of $Y_{bus}$ equals the number of contingency cases (number of post-contingency matrices) while $PQ$ requires only once.
  
\begin{equation}
  {S_{bus}=V}_{bus}\cdot{(Y_{bus}\cdot V_{bus})}^\ast
  \label{eq:calc_nodal_power}
\end{equation}

As shown in \cref{fig:form_nr}a and given in \cref{eq:calc_nodal_power}, \ac{PF} problem is defined as finding the unknown complex voltage $V_{bus}$, which minimized the power injection mismatch between the given complex bus injection $S_{bus}$ and the one calculated with $V_{bus}$ \cite{Grainger.1994}. With different bus types, different values are given as input. For slack bus the complex voltage is given, while for PV bus the constant voltage magnitude and active power and for the PQ bus with active and reactive power is predefined (shown in \cref{fig:form_nr}a). The complex voltage is represented in polar notation ($ \angle V, |V|$) and complex power injection $S_{bus}$ is given in cartesian form (P, Q). For \ac{NR} algorithm, the following processes are performed iteratively until maximum allowed iteration number reached or the maximum of the power injection mismatch drops below the tolerance (typically $10^{-8} p.u.$\cite{Zimmerman.2011}).
    \begin{enumerate}
      \item Calculate \ac{NPM} $ \Delta P_{V}, \Delta Q_{V} $ with $V_{bus}$
      \item Create/Update \ac{JM} $J$ with $V_{bus}$
      \item Solve the linear system
      \item Update $V_{bus}$ with $\Delta \angle V_{pv,pq}, \Delta |V_{pq}| $
    \end{enumerate}
In each iteration, the convergence (power injection mismatch) is checked with the \ac{NPM} calculation. \cref{sec:batched_linear_solver} explained the details of solving linear system step with LU-refactorization.

\subsection{Batched operation on sparse matrix}
  \begin{figure*}[hbt!]
    \centering
    \includegraphics[width=0.9\linewidth]{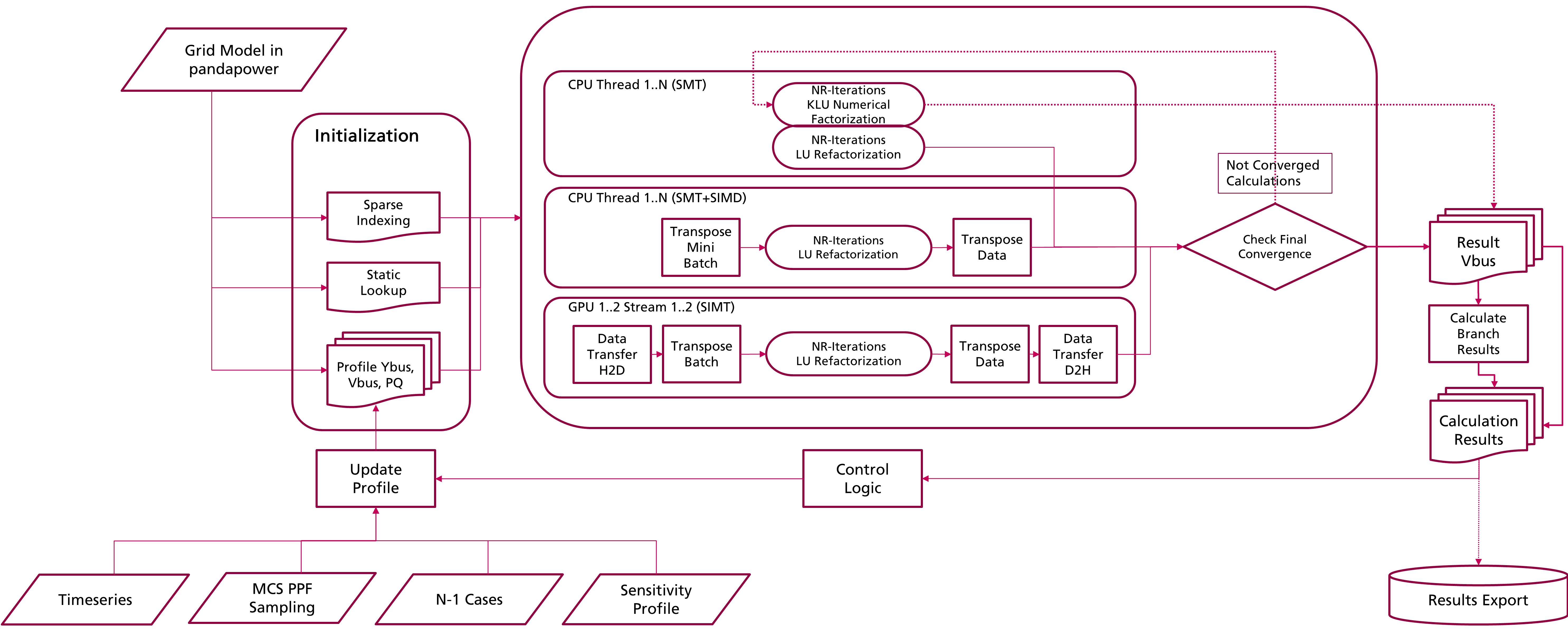}
    \caption{System overview of parallel \ac{PF} solver.}
    \label{fig_process_overview}
  \end{figure*}
  
  The utilization of the sparse matrix is essential to reduce the required memory space and avoid useless arithmetic operations on zeros, due to the high sparsity rate in $Y_{bus}, J$. Since the sparse matrices $Y_{bus}$, $J$ share the same sparsity pattern, when iterating over the elements, it is possible to broadcast operations performed on these matrices among tasks within batch. The storage of the batched profile data in memory is shown in \cref{fig:form_nr}d). The batched profile data in memory is aligned on the same element within the batch (mini-batch) and stored in contiguous memory address instead of within each task vector. On GPU, this property guarantees automatically coalesced memory access and avoid the threads divergence within one warp \cite{Zhou.2017}, as long as the batch has a size of multiply of the default warp size (32 for CUDA). With SMT+SIMD on CPU with mini-batch, it makes full utilization of the \ac{FP} capability of the instruction set. A mini-batch of size 4 is used for the efficient usage of AVX2-instruction set. 

  In order to achieve an early saturation of the GPU performance or memory bandwidth depends on the property of the kernel. Further actions to increase the saturation is required, as explained below.

  \subsubsection{Calculate nodal power injection mismatch}

  \begin{algorithm}[hbt!]
    \caption{Batched calculation of NPM}
    \label{ag:calc_error}
    \begin{algorithmic}[1]
      \STATE \# Do in parallel on CUDA
      \STATE \# Find taskId (tId) with thread indexing
      
      \STATE $row = ThreadBlock.Idy$
      \STATE $batchId = ThreadBlock.Idx$
      \STATE $tIdinBatch = Thread.Idx$
      \STATE $tId = batchId \cdot ThreadBlock.DimX + tIdinBatch$

      \STATE \# Initialize P, Q mismatch vector
      \STATE $ \Delta P_{V}(row, tId) = - P_0(row, tId)$
      \STATE $ \Delta Q_{V}(row, tId) = - Q_0(row, tId)$
  
      \STATE \# Calculate nodal power injection 
      \FOR {$dId$ in $RowPtr(row:row+1)$}
        \STATE {$ col = ColIx(dataIx) $}
        \STATE $ |S| = |V(row, tId)| \cdot |Y(dId, tId)| \cdot |V(col, tId)|$
        \STATE $ \angle S = \angle V(row, tId) - \angle Y(dId, tId) - \angle V(col, tId)$
        \STATE \# Update P, Q mismatch
        \IF {row in $bus_{pv,pq}$}
          \STATE $ \Delta P_{V}(row, tId) += |S|\cdot cos(\angle S)$
        \ENDIF
        \IF {row in $bus_{pq}$}
          \STATE $ \Delta Q_{V}(row, tId) += |S|\cdot sin(\angle S)$
        \ENDIF
      \ENDFOR

    \end{algorithmic}
  \end{algorithm}
  
  Calculation of \ac{NPM} with $V_{bus}$ is given in \cref{eq:calc_nodal_power}. The process can be performed with iterating over the element in the $Y_{bus}$ once. Due to the task independency between rows in $Y_{bus}$, extra row-level parallelism can be achieved with one CUDA thread responsible for a row of one task. \cref{ag:calc_error} shows the proposed process on GPU, which is a computing-bounding task, the extra row-level parallelization improves the kernel saturation on GPU. On CPU platform, with SMT parallelization the $taskId$ is given as an input. For SMT+SIMD parallelization the arithmetic operation is automatically broadcasted to all the task within the mini-batch. 
    
  \subsubsection{Update Jacobian Matrix}
  Similar to \cref{ag:calc_error}, update of $J$ matrix is a computing-bounding task. Due to the independency between rows, the process can be extended with row-level parallelization for a better saturation.

  In the first step, only the diagonal elements are calculated with help of the $DiagPtr$ as shown in \cref{fig:form_nr}b. On the second step, the non-diagonal elements are calculated by iterating over the $Y_{bus}$ once, during which the diagonal elements are updated. With the $d\{P,Q\}/d\{\angle V,|V|\}$ matrix consisting of four sub-matrices available, it can be subset into $J$ matrix (without P,Q for slack and Q for PV bus) or direct into permuted form $A$ with static lookup.

\subsection{System overview of parallel PF solver and its application}

  \cref{fig_process_overview} shows the method of integrating the aforementioned parallel \ac{PF} solver with pandapower. The initialization step is carried out with the pandapower in python environment including the initialization of the sparse indexing of $Y_{bus}$ and $J$ as well as the static lookup for updating sparse matrix and the profile data. With unified initialization step, the \ac{PF}s can be solved with the three types (\ac{SMT}, \ac{SMT}+\ac{SIMD}, GPU SIMT) of parallel \ac{PF} solver. The \ac{SMT}+\ac{SIMD} requires the extra transpose to mini-batch step, while the GPU version requires the memory transaction between device and host and the transpose to batch step on device. The resulted power injection mismatch is checked on CPU, if any \ac{PF} task cannot converge due to numerical instability during LU refactorization, this task will be given a second chance to get fixed with the KLU numerical factorization. The branch flow is calculated in parallel after the final convergence check with the resulted $V_{bus}$. 
  
  The proposed method, as indicated by its parallel nature of performing task-level \acp{PF} under the circumstances that $Y_{bus}$ and $J$ remains the same sparsity pattern, can profit at largest, when no dependency needs to be considered between \acp{PF}. This kind of use cases include \ac{PPF}, N-1 analysis, stationary time-series analysis without switch configuration change and some training processes in machine learning. For use case like quasi-static time-series simulation in distribution grid, in which the discrete actions such as \ac{OLTC} tap position, status of shunt compensator rely on the status continuity but has no impact on the sparsity pattern. This type of simulation can still be accelerated by the proposed method, with an external control logic to update the profile with the persistance of the continuity of discrete variable. However, the control logic might lead to extra loops to correct the excessive status changes, which impacts negatively on the computational efficiency. For simulations with topology changes, such as switch position or status change, the extra initialization (costly, as shown in \cref{tab:final_benchmarking}) needs to be performed for each new topology. When only limited variations of grid configuration needs to be considered, the proposed method can still accelerate the overall simulation.

  The proposed approach is realized with C++/CUDA with an interface to Python with Cython 0.29.21\cite{Behnel.2011}. The programming language Python has convenient tool for data integration and processing e.g. Pandas (High-Level)\cite{McKinney.2010} and efficient numerical operations with Numpy \cite{Harris.2020}. The underlying data stored with Numpy array stored in C format can be shared and ported directly to the parallel \ac{PF} solver.

%% file: batched_linear_solver.tex
The solving of linear system is a time-consuming step in solving \ac{NR} \ac{PF}, as it is in the actual pandapower implementation\cite{Schafer.}. The mathematical formulation of solving linear system step in \ac{NR} process is given in \cref{eqn:solve_linear_system}. The $b$ is the active and reactive power mismatch with the given $V$, the result $x$ is used to update the $V$. 

\begin{align}
  \label{eqn:solve_linear_system}
  \begin{split}
    & J_{1..n} \cdot x_{1..n} = b_{1..n}
  \\
  & x = [\Delta \angle V_{pv,pq}, \Delta |V_{pq}| ] \\
  & b = [\Delta P_{V}, \Delta Q_{V}]
  \end{split}
\end{align}

In this work, the existing process of solving linear system is analyzed and an optimized algorithm for batched linear system solver is proposed.

\subsection{Process of solving with direct LU solver}

\begin{figure*}[hbt!]
  \centering
  \includegraphics[width=0.8\linewidth]{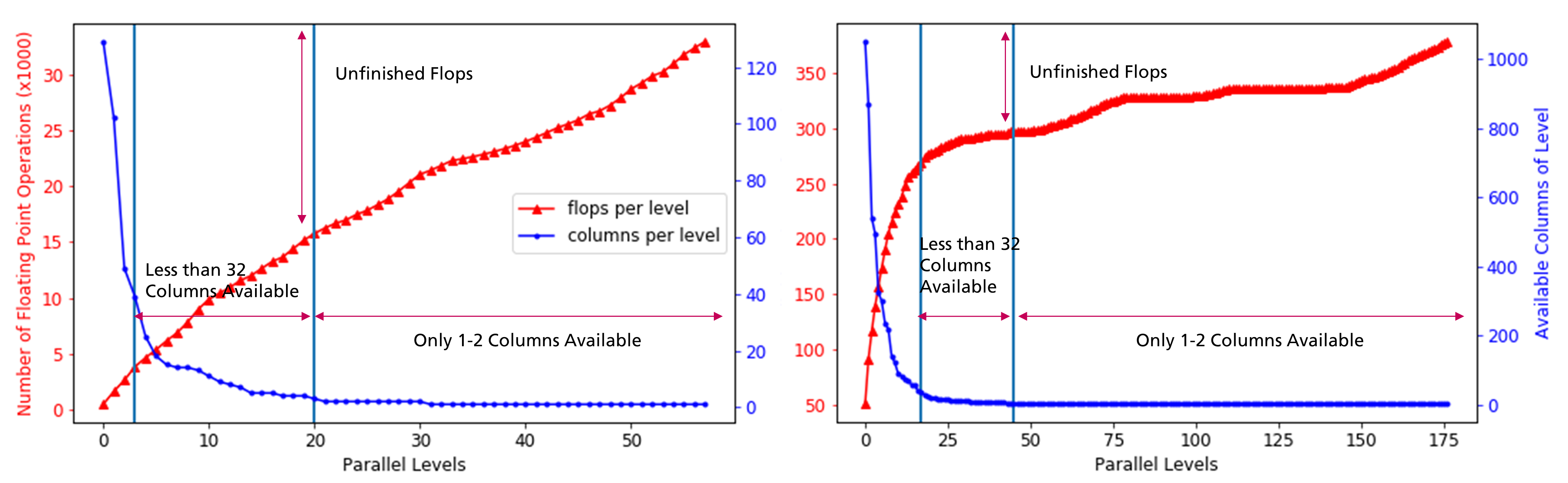}
  \caption{LU Parallel Levels and \ac{FP}s (left: IEEE case300, right: case2869pegase).}
  \label{fig:lu_col_level}
\end{figure*}

In solving linear system with LU factorization, the original linear system is pre-ordered (pre-scaled), factorized into lower-triangular matrix $L$ and upper-triangular matrix $U$ and finally solved with \ac{FS}-\ac{BS}, as it is in the modern implementations \cite{Davis.2010,Li.2005,Peng.01.08.2019,Chen.2015}. KLU is considered one of the fastest single-thread linear solver for \ac{JM}\cite{Razik.,Su.2020,Zhou.2017,Zhou.2018}, which is used as the reference for the proposed method in this work. The KLU solver can be easily extended for task-level parallelization with OpenMP. Following theoretical introduction focuses on the implementation of KLU and gives hints on the optimization for solving many \ac{PF}s with same $Y_{bus}$ sparsity pattern.

In the pre-ordering phase, the original $J$ is permuted in order to reduce the required \ac{FP}s by reducing the fill-ins. Multiple heuristic methods are available for the pre-ordering, which gives different performance related to the matrix pattern. In this work, the pre-ordering method (AMD \cite{Amestoy.2004}) available in KLU is used, since \cite{Razik.,Huang.2018b} reported its good performance in reducing fill-ins generally on circuit simulation and \ac{PF} analysis with \ac{NR} algorithm. In other cases, such as to factorize augmented \ac{JM} to directly consider control effect from e.g. \ac{OLTC}, the work \cite{Kocar.2014} presented an extra analysis regarding the pre-ordering method. With the permutation, the original linear system $J$ is permuted into the $A$ as given in \cref{eqn:linear_system_perm}. $Perm_{col}$, $Perm_{row}$ are the correspondent column and row permutation matrix. The numerical factorization is performed on $A$. 

\begin{align}
  \label{eqn:linear_system_perm}
  \begin{split}
    & A_{1..n} = Perm_{col} \cdot J_{1..n} \cdot Perm_{row} \\
    & A_{1..n} = L_{1..n} \cdot U_{1..n}
  \end{split}
\end{align}

In the numerical factorization of KLU, the \ac{G-P} algorithm is utilized. Additionally, partial pivoting is performed to improve the numerical stability, with which the permutation matrix $Perm_{row}$ is updated, in order to avoid the tiny pivot. This step has effect on the final sparsity pattern on $A$.

Refactorization is a much faster process, which reduced the computational overhead by presuming the numerical stability of the permutation matrix $Perm_{row}$, $Perm_{col}$. For \ac{NR} iterations in solving \ac{PF} and circuit simulation, refactorization is preferred \cite{Chen.2015}. The refactorization mode is also supported by KLU.

The pattern of $A$, $L$ and $U$ remains unchanged for all tasks when solving many \ac{PF}s with same $Y_{bus}$ sparsity pattern. The pre-ordering only need to be performed once at the beginning \cite{Zhou.2017b}. Based on this property, a static lookup can be created with the final permutation matrices $Perm_{row}$, $Perm_{col}$ found. Since in \ac{G-P} algorithm \ac{CCS} matrix is required, which is efficient for iterating over columns. A static lookup can direct convert the original \ac{CRS} \ac{JM} or $d\{P,Q\}/d\{\angle V,|V|\}$ matrix into the permuted $A$ into \ac{CCS} format.

\subsection{CPU LU Refactorization}

\begin{algorithm}[hbt!]
  \caption{Sparse \ac{G-P} refactorization algorithm with column working space}
  \label{ag:lu_cpu}
  \begin{algorithmic}[1]
  \FOR {tId in 0:$N_{task}$}
    \STATE \# do in parallel on CPU
    \FOR {col in 0:$N_{col}$}
      \STATE {\# Copy column to working space}
      \STATE $ x = A(:, col, tId) $
      \FOR {row in URowIx(:, col)}
        \STATE {\# Sparse VMAD on column working space}
        \STATE $ x(row+1:) -= x(row) \cdot L(:, row, tId) $
      \ENDFOR

      \STATE \# Normalize L and update LU
      \STATE $ U(:, col, tId) = x(:col) $
      \STATE $ L(:, col, tId) = x(col+1:) / x(col) $
    \ENDFOR
  \ENDFOR
  \end{algorithmic}
\end{algorithm}

\cref{ag:lu_cpu} gives the SMT LU refactorization algorithm. Sparse \ac{G-P}\cite{JohnR.Gilbert.09.1986} refactorization on CPU with column working space is implemented. The working space has the size of the dimension of $A$ for SMT. With the column working space $x$, only the copy of non-zero values of $A$ is required for the sparse \ac{VMAD}.

For SMT+SIMD, working space with the width of the mini-batch size is required. By replacing $tId$ with the ID of mini-batch, the copy, \ac{VMAD} and normalization can be extended with SIMD instructions, so that the operation is broadcasted within mini-batch.

Since task-level parallelization can fully saturate all physical cores of CPU with \ac{SMT} and \ac{SMT}+\ac{SIMD}, column-level parallelization as proposed in \cite{Chen.2013} is not needed.

\subsection{GPU batched LU Refactorization and FS-BS}

\begin{figure}[hbt!]
  \centering
  \includegraphics[width=0.45\textwidth]{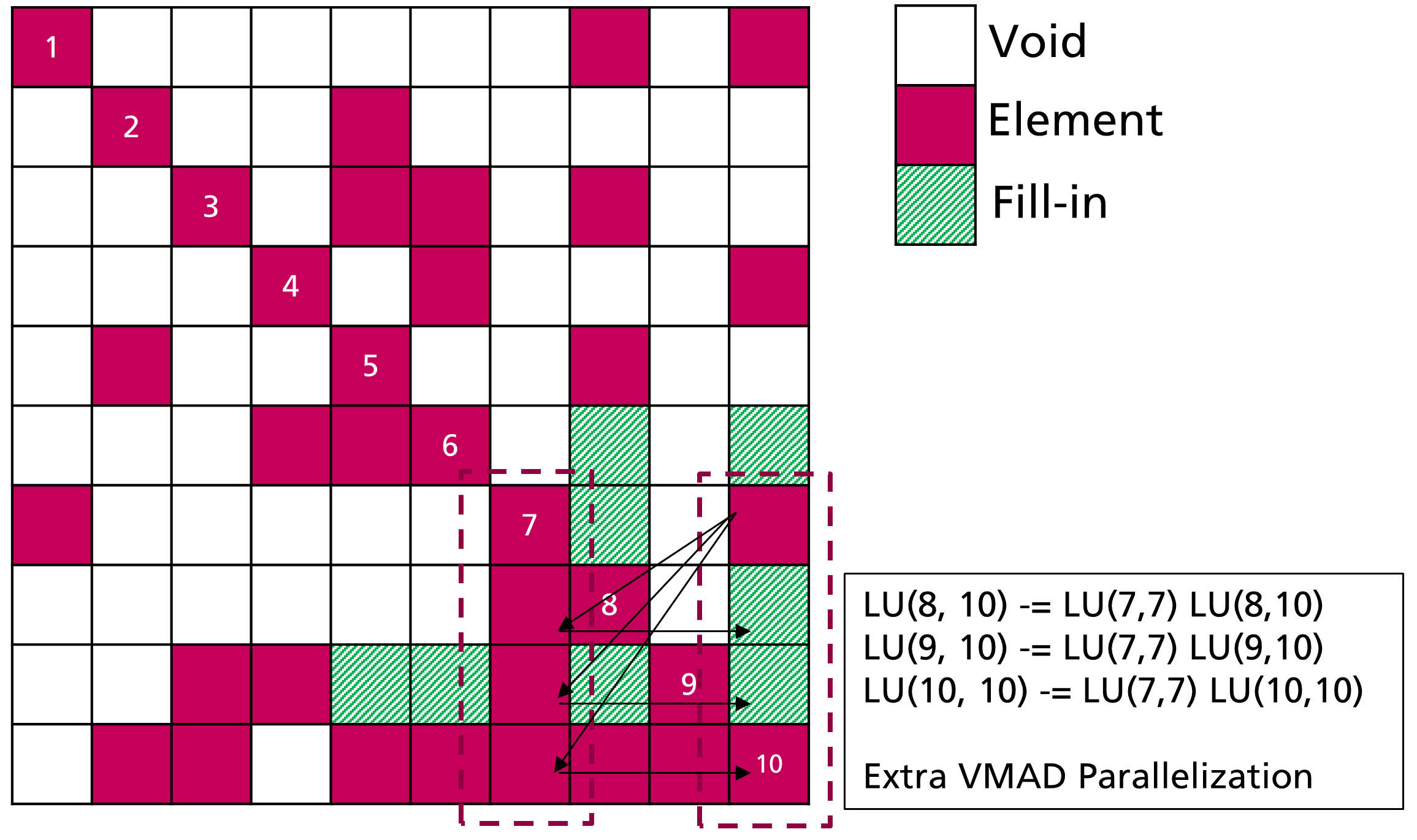}
  \caption{Example LU matrix for LU Refactorization.}
  \label{fig:lu_a_fillin}
\end{figure}

\begin{figure}[hbt!]
  \centering
  \includegraphics[width=0.45\textwidth]{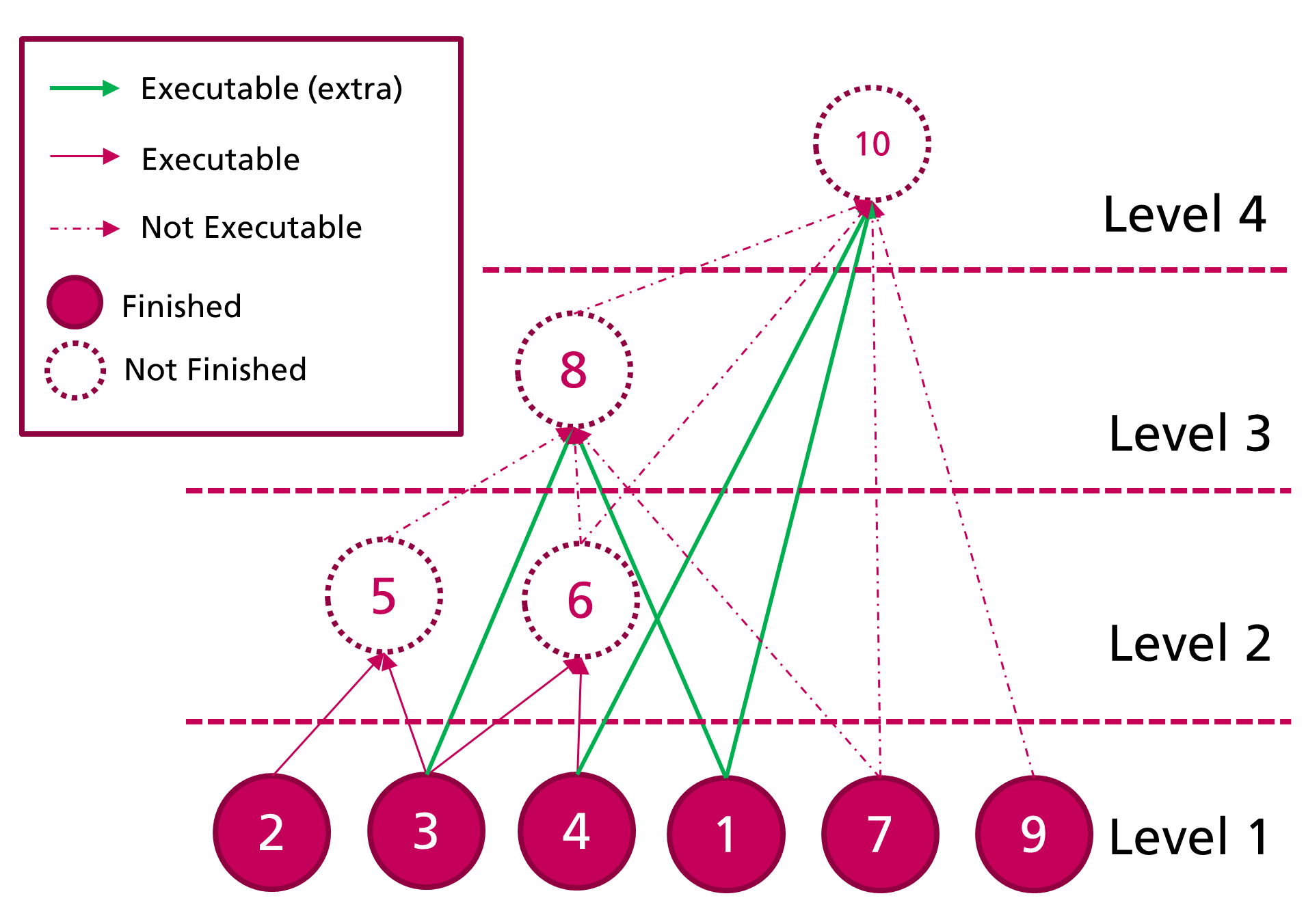}
  \caption{Example scheduling graph for LU Refactorization.}
  \label{fig:schedule_a_fillin}
\end{figure}

\begin{figure}[hbt!]
  \centering
  \includegraphics[width=0.45\textwidth]{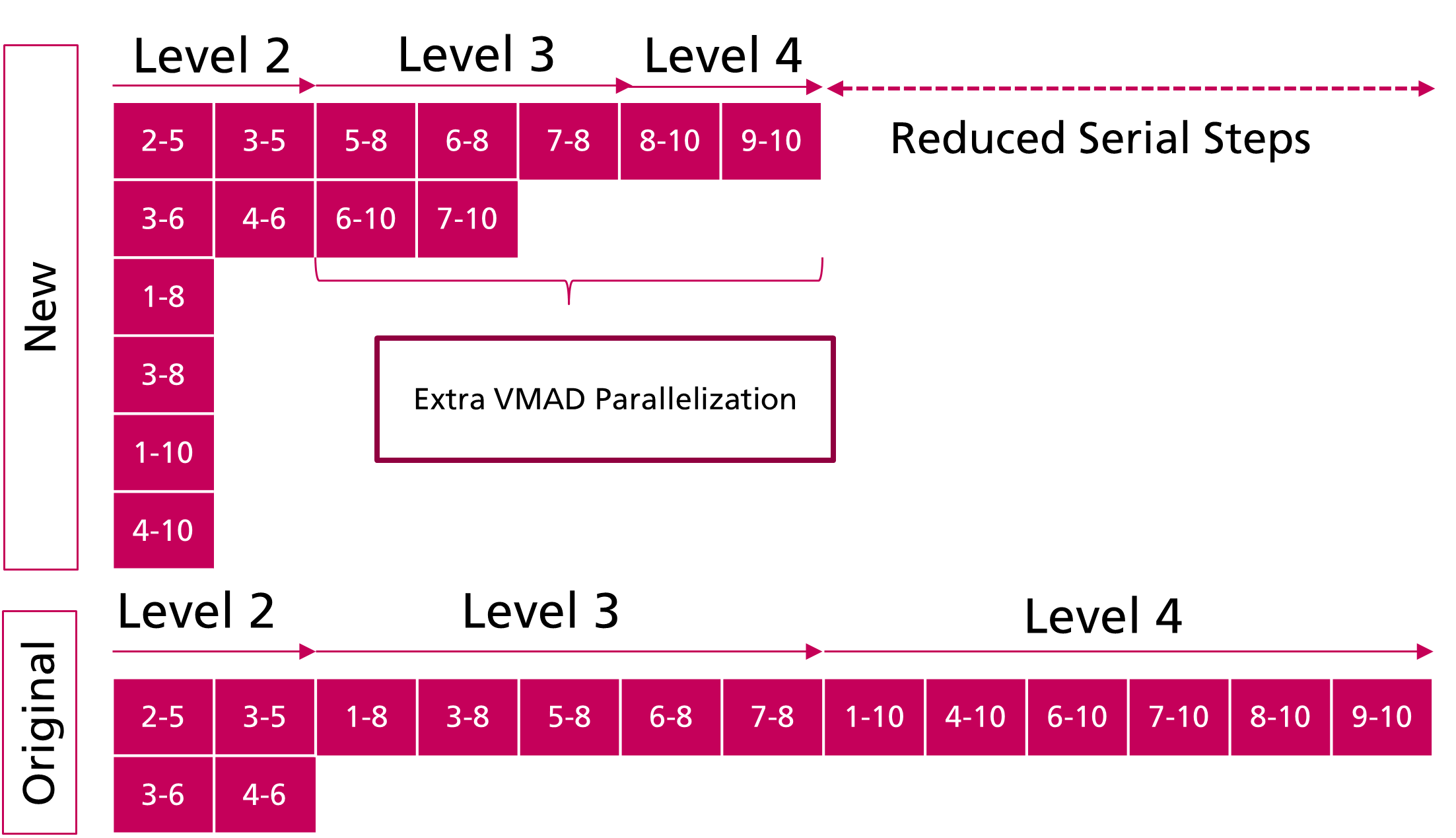}
  \caption{Task scheduling on time line for LU Refactorization.}
  \label{fig:timeline_schedule_a_fillin}
\end{figure}

\subsubsection{Theoretical analysis}
Recent effort on LU factorization with GPU is presented in \cite{Chen.2015, Peng.01.08.2019}, both works focus on accelerating single LU factorization on GPU, especially for large scaling matrix. For batched LU factorization, \cite{Zhou.2017} presents a batched LU-factorization scheme with batch and column-level parallelization. It is essential to saturate the GPU memory bandwidth for LU refactorization and \ac{FS}-\ac{BS} to achieve good performance on GPU. Due to the underlying data dependency, not all columns can be factorized simultaneously. A \ac{DAG} is used to describe and schedule the columns which can be finished at the same time \cite{Nechma.2015}. The vertices represent the columns and the edges represent the operation between two columns. An example matrix and its \ac{DAG} is shown in \cref{fig:lu_a_fillin} and \cref{fig:schedule_a_fillin} respectively. 

 Considering the \ac{G-P} algorithm with column-level parallelization proposed in \cite{Zhou.2018}, as the example of two standard grids shown in \cref{fig:lu_col_level}, at the beginning with high number of available columns per level, the GPU resource can be easily saturated even with small batch size. However, significant amount of \ac{FP}s are located in the serial region, where only few columns can be factorized at the same time. Using NVIDIA GTX1080 graphics card as an example, the GPU has 20 \ac{SM}s with each \ac{SM} has 128 \ac{SP}s. Assuming the batch size is 512, which equals to less than 1 active warps per SM, which means 75\% of the \ac{SP} remains idle when no extra parallelization is applied. Consequently, the memory-bandwidth cannot be efficiently utilized.
 
 Inspired by the work \cite{Chen.2015}, our work improves the batched LU-refactorization with fine-grained parallelization for serial levels, which improves the saturation of hardware significantly with small batch size. This improvement make greater sense on the future applications with even more computational resources and higher memory-bandwidth available on new generations of GPUs.

\subsubsection{Proposed algorithm on GPU}

\begin{algorithm}[hbt!]
  \caption{Multi-stage sparse \ac{G-P} algorithm on CUDA}
  \label{ag:lu_without_x}
  \begin{algorithmic}[1]
    \STATE \# DEFINE VMAD parallelization width as N
    \STATE \# Do in parallel on CUDA
    \STATE \# Find task (tId) with thread indexing
    \STATE $batchId = ThreadBlock.Idx$
    \STATE $tIdinBatch = Thread.Idx$
    \STATE $tId = batchId \cdot ThreadBlock.DimX + tIdinBatch$

    \STATE \# IF STAGE 1
    \STATE $col = AvailableColsinLevel(ThreadBlock.Idy)$
    \STATE \# IF STAGE 2,3
    \STATE $col = LeftoverCols(ThreadBlock.Idy)$
  
    \FOR {row in URowIx(:, col)}
      
      \STATE \# IF STAGE 2,3
      \IF {row not in FinishedCols}
        \STATE {break}
      \ENDIF

      \STATE \# IF STAGE 2,3
      \IF {row in FinishedRowsOfCol(col)}
        \STATE \# Skip finished rows
        \STATE {continue}
      \ENDIF

      \STATE {\# Sparse VMAD with direct indexing}
      \STATE {\# IF STAGE 2 element-wise iteration in vector}
      \STATE {\# IF STAGE 3 N-elements-wise iteration in vector}
      \STATE $ LU(row+1:, col, tId) -= $
      \STATE {$ LU(row, col, tId) \cdot LU(row+1:, row, tId) $}
    \ENDFOR

    \STATE \# Check column is finished
    \IF {$row == URowIx(-1, col)$}
      \STATE \# Set flag on finished columns
      \STATE UpdateFinishedColStatus(col)
      \STATE \# Normalize L
      \STATE {\# IF STAGE 2 element-wise iteration in vector}
      \STATE {\# IF STAGE 3 N-elements-wise iteration in vector}
      \STATE $ LU(row+1:,col, tId) /= LU(row, col, tId) $
    \ELSE
      \STATE \# IF STAGE 2,3
      \STATE \# Memorize the processed row of the column
      \STATE UpdateFinishedRowsOfCol(row, col)
    \ENDIF

  \end{algorithmic}
\end{algorithm}

A three-stage batched \ac{G-P} algorithm is proposed for the GPU refactorization, which is given in \cref{ag:lu_without_x}. $LU$ denotes the working space for LU refactorization ($L+U-I$) including all the fill-ins predefined. The values in A needs to be copied into $LU$ with the fill-ins initialized as 0. 

In stage 1, with large number of columns available in each level, the columns are fully factorized in parallel. Each CUDA thread is responsible for one column in one task. In stage 2, besides the columns, which can be fully factorized, the viable operations for all the unfinished columns can be executed simultaneously (shown in \cref{fig:schedule_a_fillin} as green lines). In stage 3, the extra \ac{VMAD} parallelization as shown in \cref{fig:lu_a_fillin} could be applied with extra CUDA threads scheduled within the same CUDA $threadBlock$, since only threads within one $threadBlock$ can be explicitly synchronized. The extra \ac{VMAD} parallelization could use width of e.g. 4. In stage 2 and 3, on each level some columns can only be partially factorized, the already processed rows of each column are memorized, the finished columns of each level will be directly normalized.

Since the available columns in each parallel levels is known after the symbolic analysis step. As indicated in \cref{fig:lu_col_level}, the start of level 2 and level 3 is solely related to the available columns in each level. This parameter is related to the hardware. As manually tuned in our tests, to achieve good performance, the stage 2 can start when less than 32 columns are available in level and stage 3 starts when only 1 column is available.

For example matrix $LU$ shown in \cref{fig:lu_a_fillin}, the level one in \cref{fig:schedule_a_fillin} corresponds to the stage 1 of \cref{ag:lu_without_x}. After the first level is finished, tasks which belong to the columns in later levels can be executed in an earlier stage, which in turn increased the saturation rate and reduced the tasks in the serial levels (see \cref{fig:timeline_schedule_a_fillin}). Level 2 corresponds to the stage 2 of the algorithm. In level 3, 4, when only one single column is available the extra \ac{VMAD} parallelization is applied according to stage 3. In this case, the row 8, 9, 10 are assigned onto the $Thread.Idy$, thus \ac{VMAD} can be performed more efficiently with suitable parallelization width instead of element-wise.

\subsubsection{GPU Forward Substitution Backward Substitution}

\begin{figure}[hbt!]
  \centering
  \includegraphics[width=0.45\textwidth]{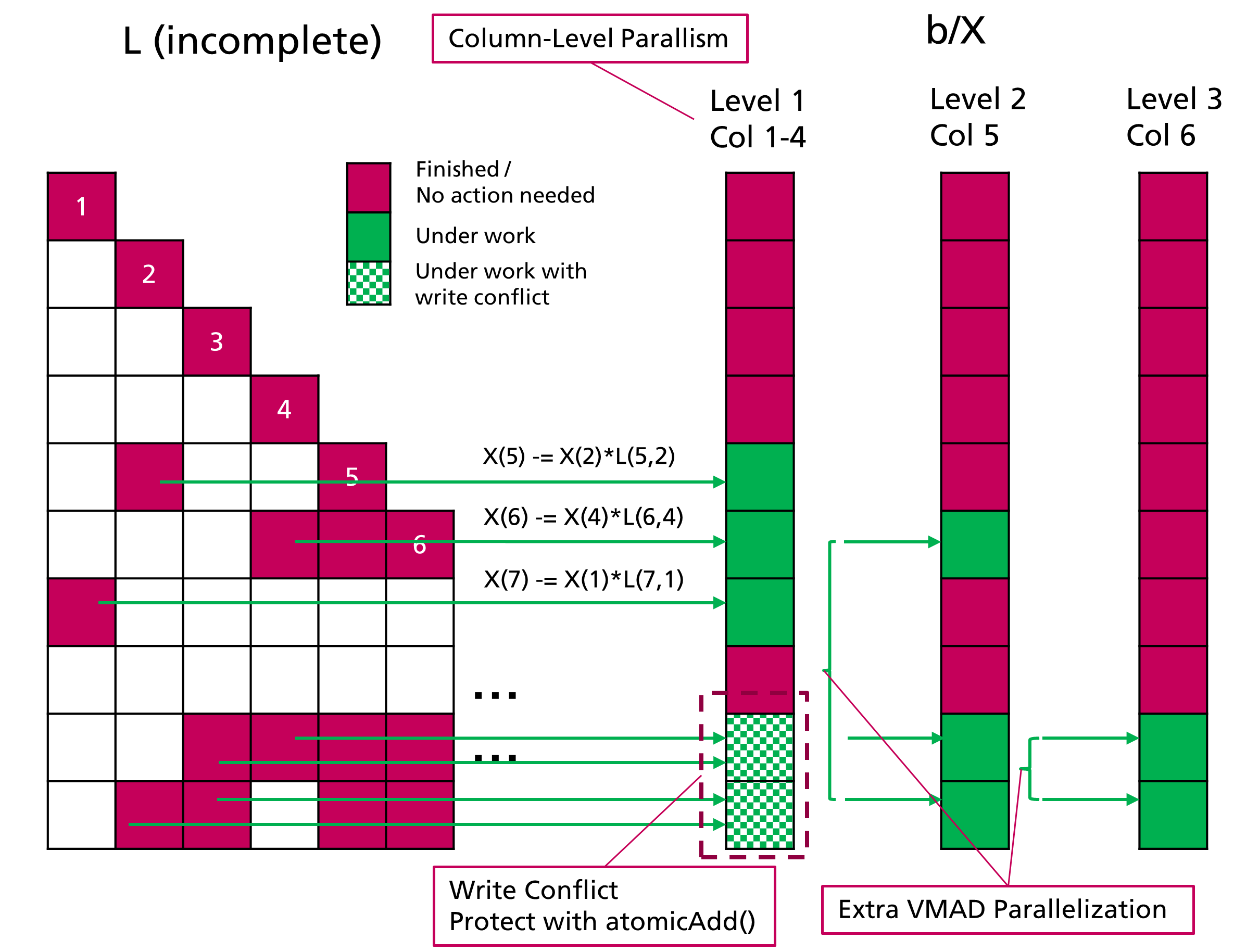}
  \caption{Parallelized \ac{FS}-\ac{BS}.}
  \label{fig:fsbs}
\end{figure}

\cref{fig:fsbs} shows the parallelization strategy of \ac{FS}-\ac{BS} with an incomplete $L$ matrix. After the applying permutation matrix to the $x$, \ac{FS} with $L$ matrix can be performed on the same working space $b/x$. A dependency graph can be constructed to guide the parallel execution of multiple columns of $L$. As soon as the pivot element (upper element) of $x$ corresponds to the column is finalized, this column can be executed to update the lower elements in $x$. When multiple threads try to update the same element, the barrier write function $atomicAdd()$ protect from writing collision. The \ac{VMAD} parallelization can be applied, when few columns are available on the same level. The same approach is applied for $U$.

%% file: case_study.tex
The case study on CPU is performed on a Windows 10 PC with Intel i7-8700 CPU (6 physical cores and \ac{HT} technology for 12 virtual cores) and 16 GB DDR4 RAM, with the prioritized Intel C++ Compiler V2020 and the Intel Math Kernel libraries, which is highly optimized for the vectorized computation. The code is compiled with O3 and forced using the AVX2 instruction set.

The case study on GPU is performed on a PC with Intel i7-8700k, 32 GB DDR4 RAM and 2x Nvidia GTX 1080-8GB GPUs. The PC is running on Ubuntu 18.04.4 LTS. The proposed approach is programmed in C++ and compiled with GCC V7.4 and CUDA V10.1 with O3. The main focus of the benchmarking is the duration of the \ac{PF} solving and each subprocess. Because the kernel execution in CUDA is initialized from CPU and executed on GPU asynchronously, to record the duration of each kernel, the test is performed with synchronization at the kernel (or multiple-kernels e.g. for LU-refactorization) exit. 

On both test platforms, double precision float number is used, due to the high numerical stability requirement during the LU refactorization. The examples are tested with small-size grid "IEEE case300" with 300 buses and mid-size grid "case2869pegase" with 2869 buses available in pandapower. Both contain meshed \ac{EHV} and \ac{HV} voltage levels, IEEE case300 contains also \ac{MV} and \ac{LV} buses.

\begin{figure*}[hbt!]
  \centering
  \includegraphics[width=0.4\linewidth]{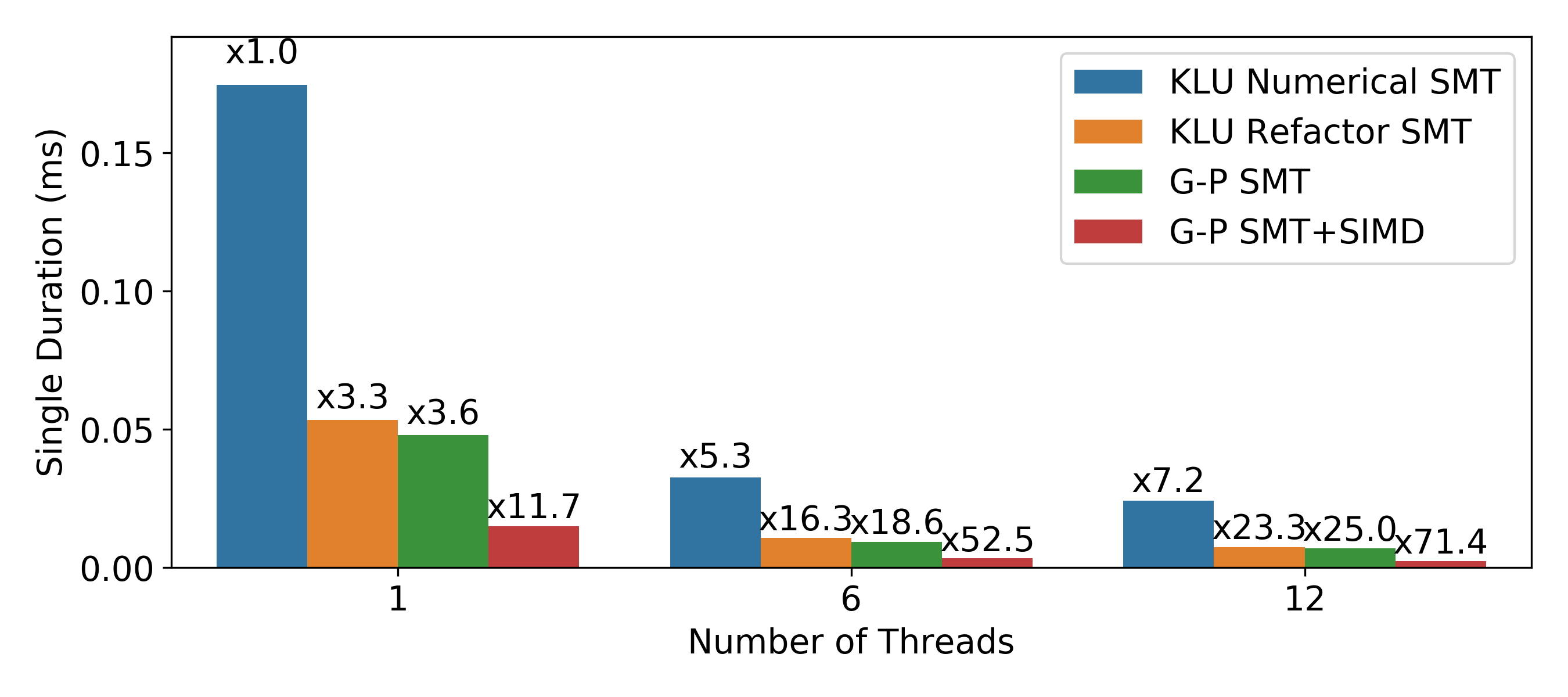}
  \includegraphics[width=0.4\linewidth]{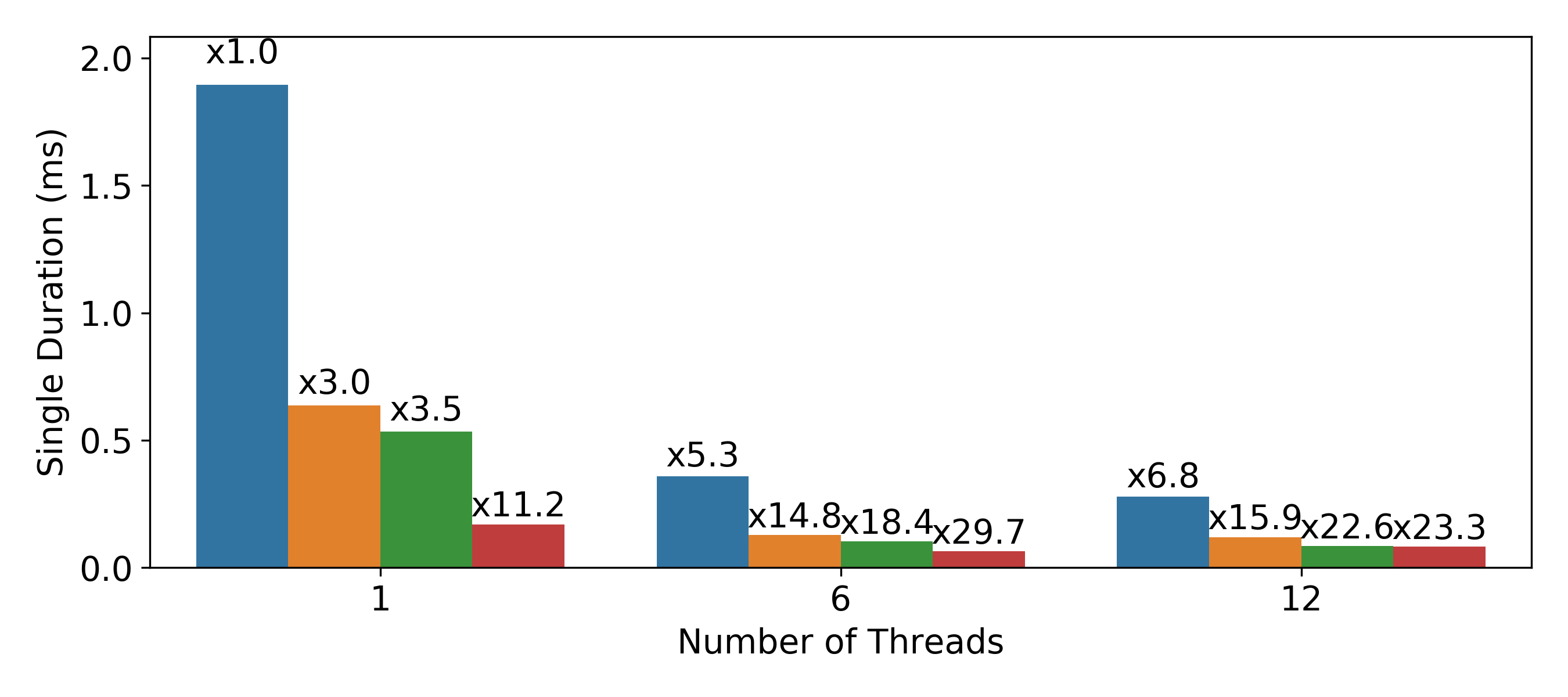}

  \caption{Batched linear solver benchmark on CPU (left: IEEE case300, right: case2869pegase).}
  \label{fig:lu_performance_cpu}
\end{figure*}

\begin{figure*}[hbt!]
  \centering
  \includegraphics[width=0.4\linewidth]{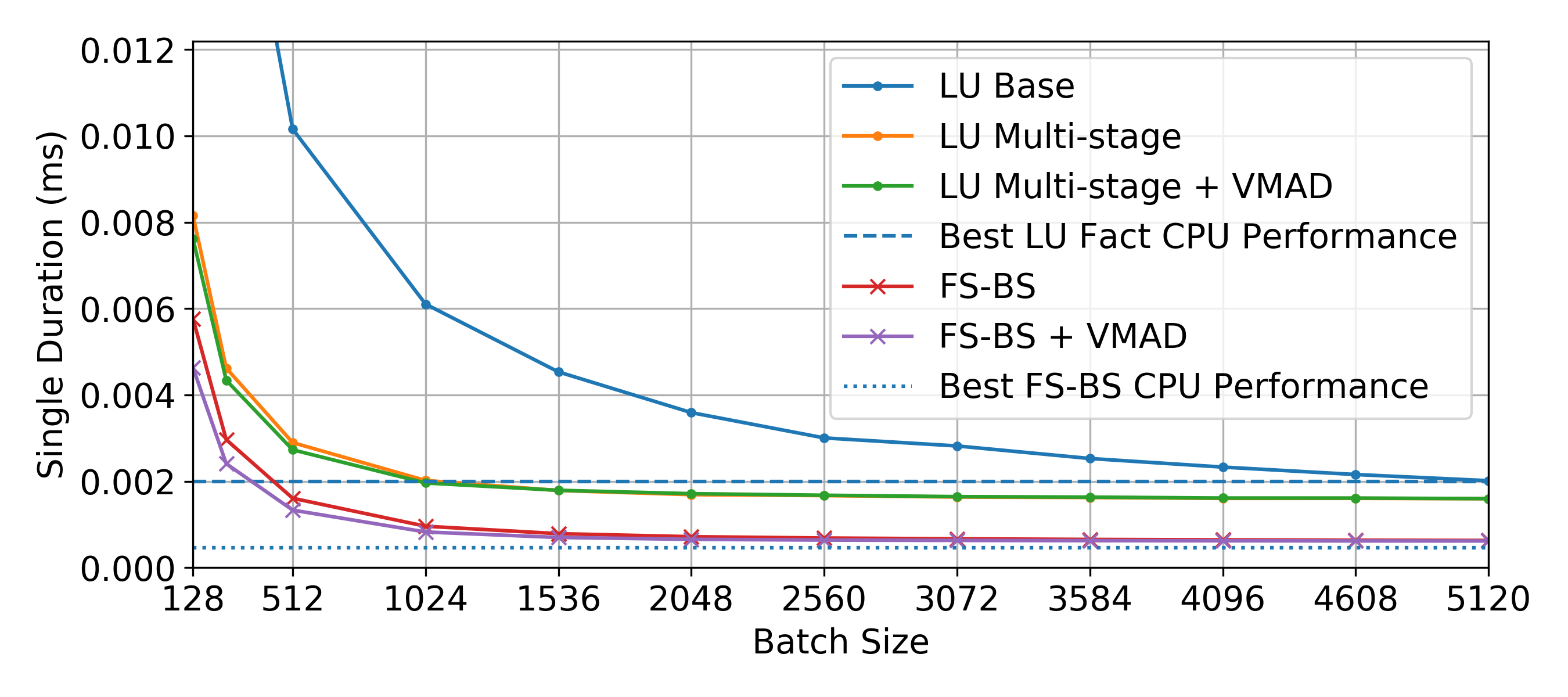}
  \includegraphics[width=0.4\linewidth]{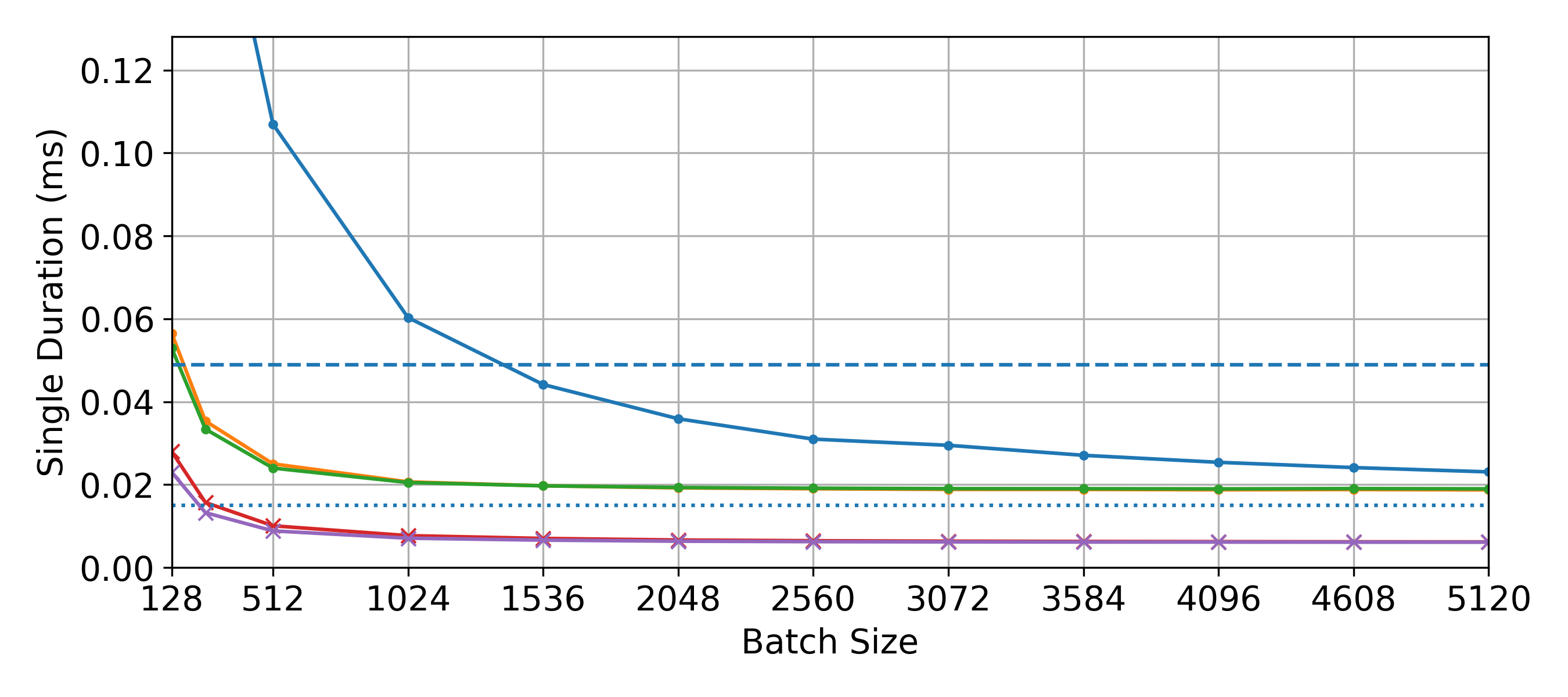}

  \caption{Batched linear solver benchmark on GPU (left: IEEE case300, right: case2869pegase).}
  \label{fig:lu_performance_gpu}
\end{figure*}

\subsection{Batched linear solver performance analysis}

This section gives a performance evaluation of the aforementioned batched linear solver on CPU and GPU and the relative performance to KLU.


\cref{fig:lu_performance_cpu} shows the performance on CPU platform. On both test grids, a performance improvement of the implemented \ac{G-P} algorithm can be observed over KLU against both numerical factorization and refactorization modes. Because it works directly on the already permuted $A$ matrix, as the KLU takes the original matrix as input. With \ac{SMT} parallelization, a performance improvement of $>$x6 can be observed for all the methods tested when increasing the number of threads, especially when below the number of physical cores (6). With further increasing the thread number, the \ac{HT} technology helps further improve the saturation of the physical core, thus improve the overall performance. With the SMT+SIMD parallelization, further speed-ups can be observed. However, a slight performance decreasing can be observed on the mid-size grid under the \ac{G-P} SMT+SIMD version, when increasing the threads number from 6 to 12, this behavior is caused by the reduced cache hitting rate, due to the large problem scaling. Overall, with the proposed method, on CPU platform, we achieved a good performance improvement of x20 - x70. The acceleration rate has a close relationship to the size of the problem (number of buses).

On the GPU platform, \cref{fig:lu_performance_gpu} shows the benchmarking result with different batch sizes, the best CPU results is used as baseline. It can be observed, that our approach of further fine-grained parallelization of the refactorization process leads to a earlier saturation of the GPU resource and achieved much better performance when the batch size is small for both grids. With the extra \ac{VMAD} parallelization with the 4x threads in stage 3, it improves the performance when the batch size is very small ($\leq 512$) for LU refactorization and \ac{FS}-\ac{BS}, slight difference can be observed when the GPU is fully saturated. When dealing with a large grid, the improvement is more significant comparing to the CPU counterpart due to the higher cache misses on CPU. 

On the GPU test with a large grid (case9241pegase with 9241 buses), the simultaneous LU refactorization requires large working space on GPU memory, which leads to the fact that only batch size up to 2048 can be supported on the test platform. By setting the batch size to 2048, the average time of one LU-refactorization with proposed multi-stage  method requires 0.138 ms while the base batch version requires 0.238 ms. The proposed method show also significant improvement in large grid.

\subsection{Performance analysis on CPU}

\begin{figure*}[hbt!]
  \centering
  \includegraphics[width=0.4\linewidth]{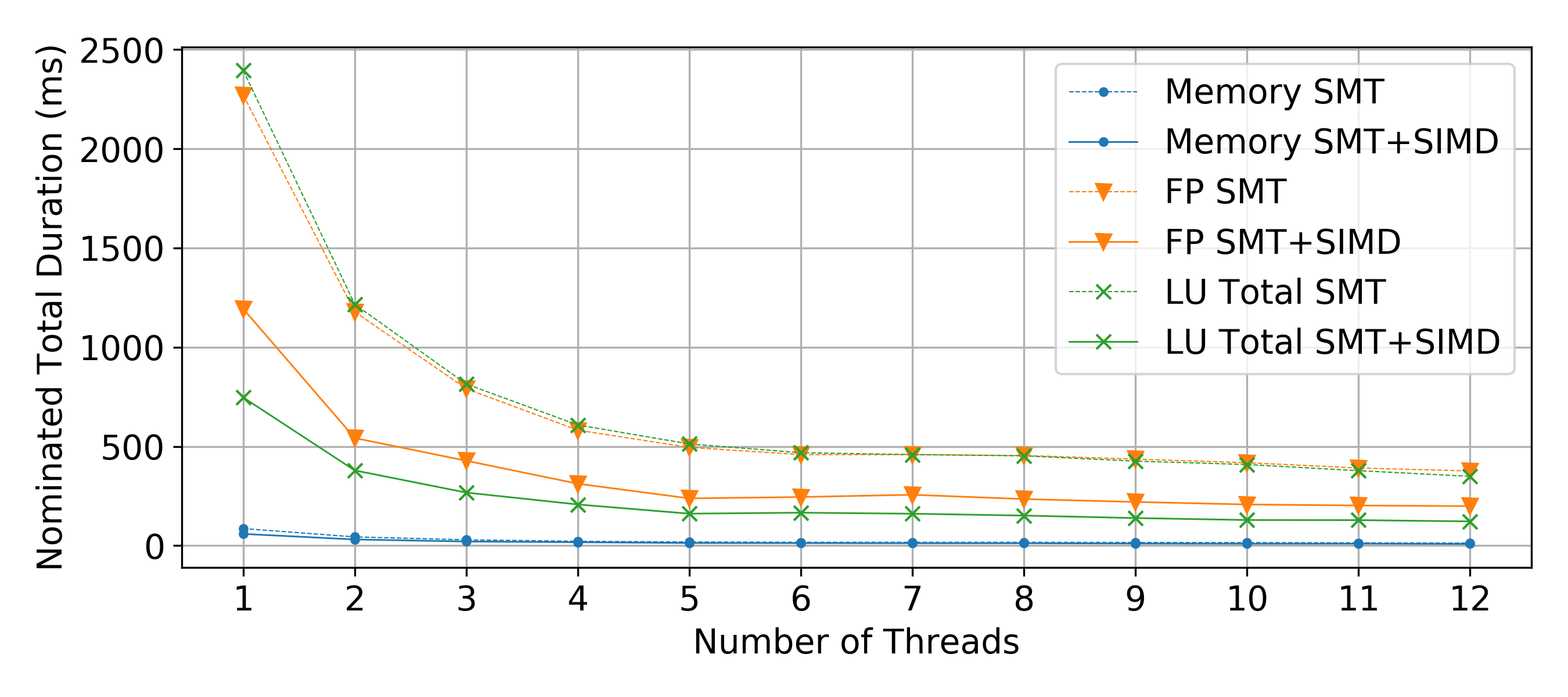}
  \includegraphics[width=0.4\linewidth]{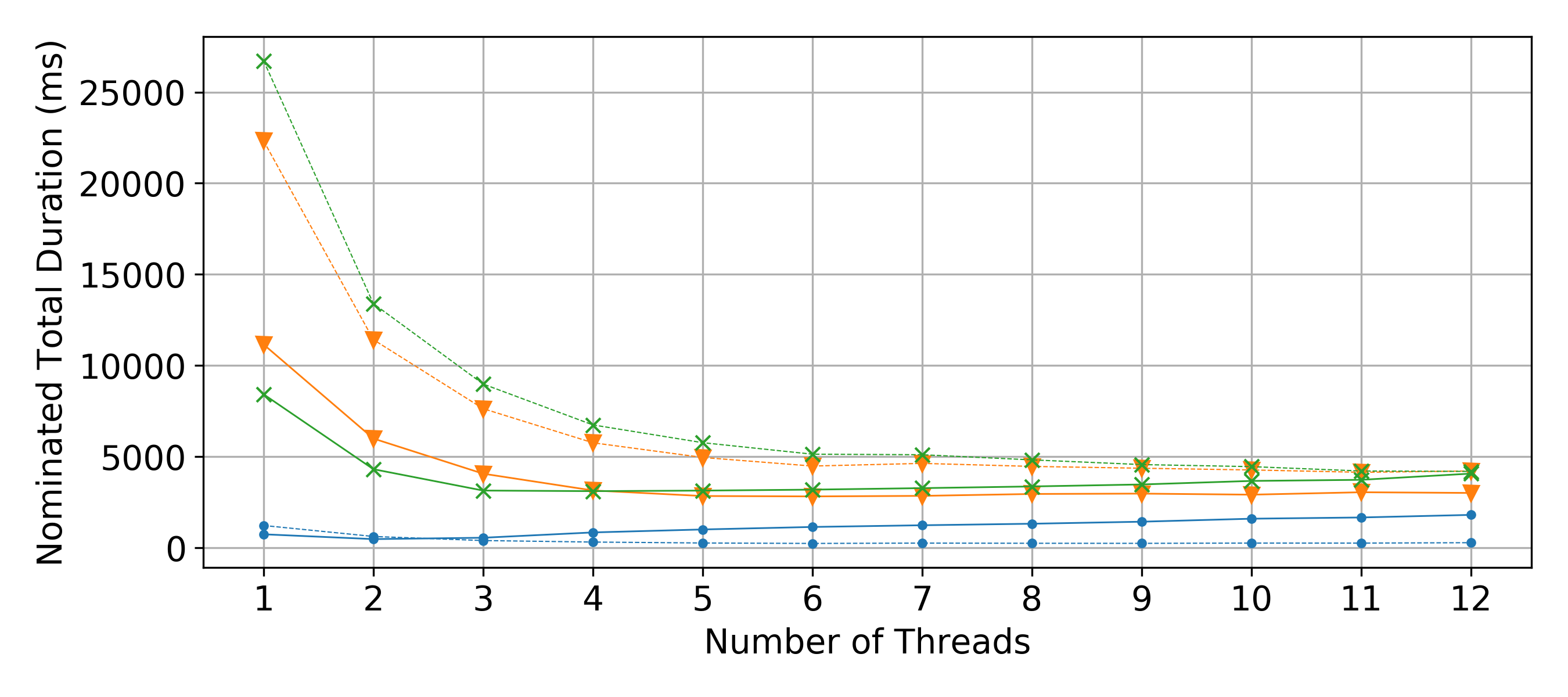}

  \caption{Function benchmark results on CPU for the performance comparison over SMT and SMT+SIMD (left: IEEE case300, right: case2869pegase).}
  \label{fig:kernel_saturation_cpu}
\end{figure*}

\begin{figure*}[hbt!]
  \centering
  \includegraphics[width=0.4\linewidth]{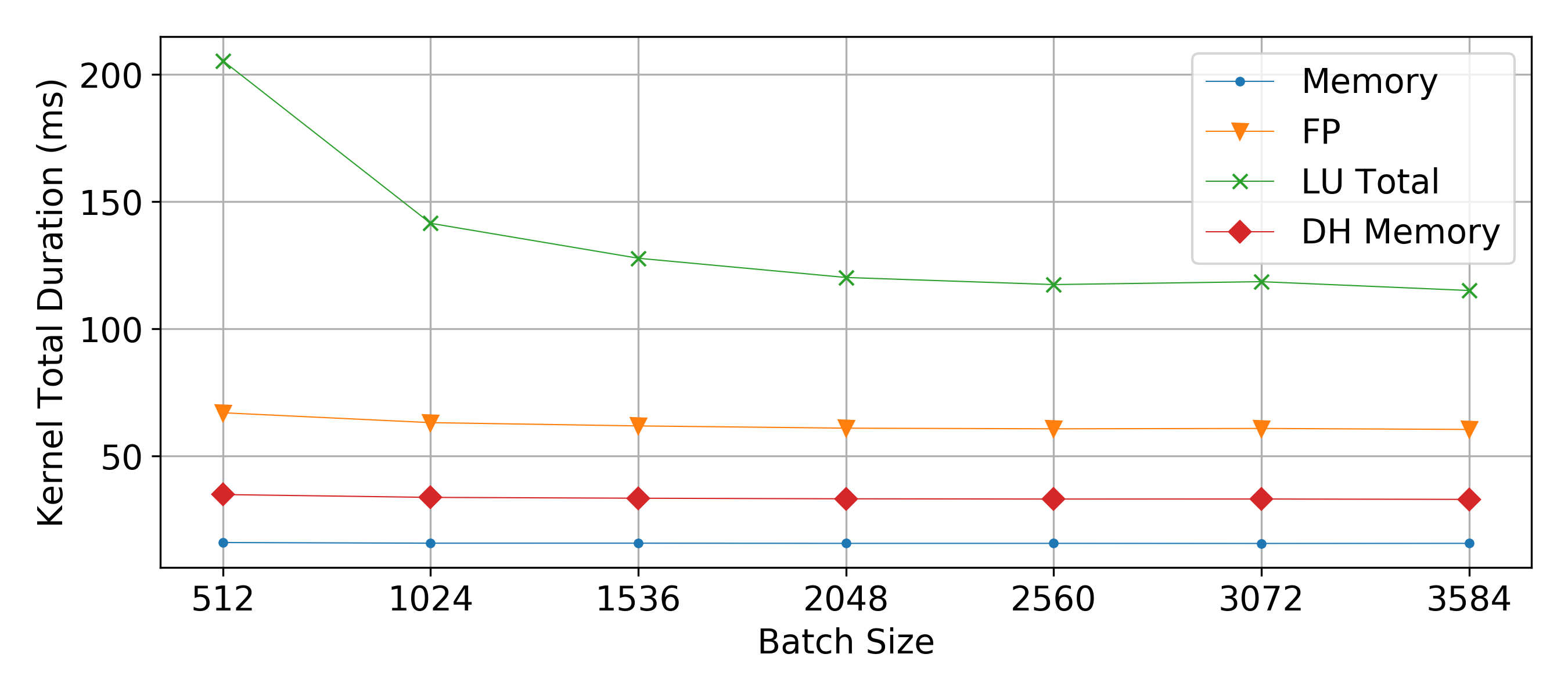}
  \includegraphics[width=0.4\linewidth]{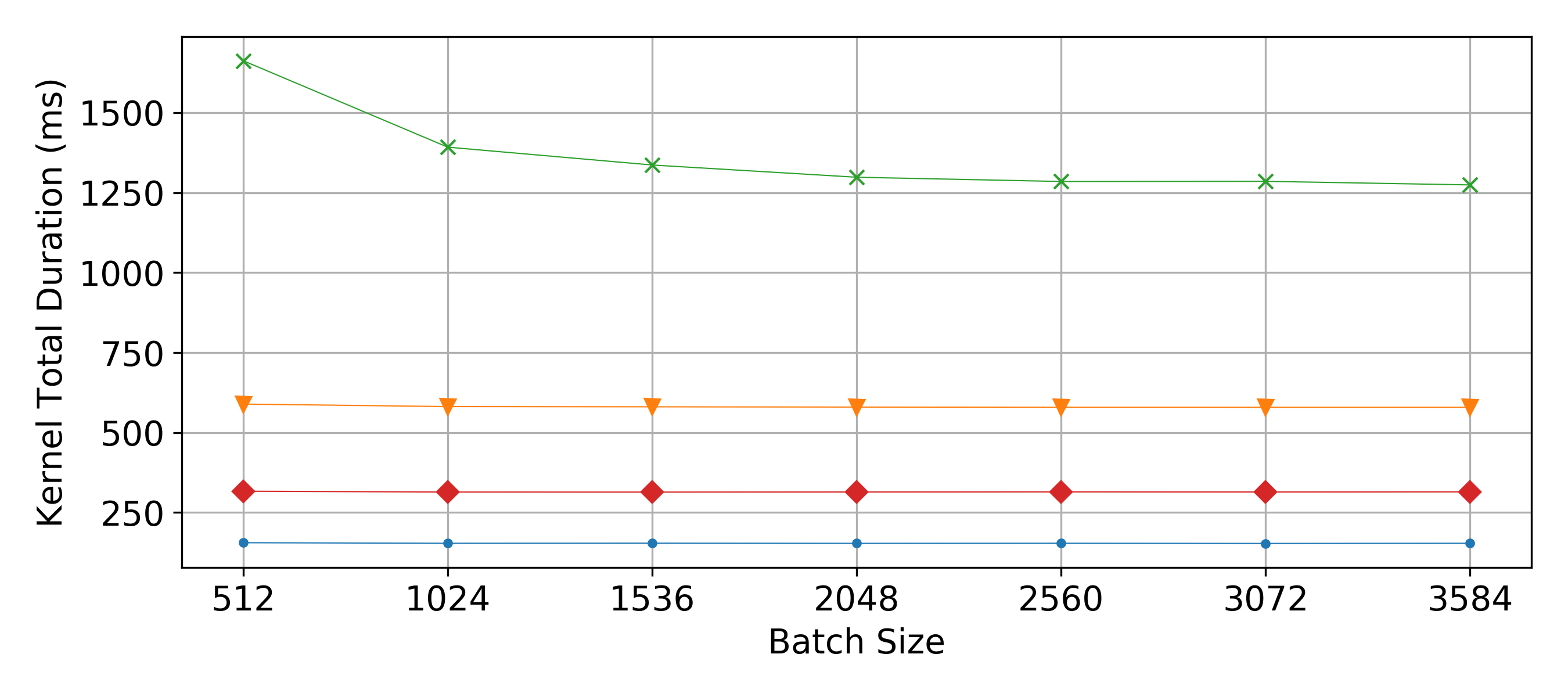}

  \caption{Function benchmark results on GPU (left: IEEE case300, right: case2869pegase).}
  \label{fig:kernel_saturation_gpu}
\end{figure*}

For the convenience of comparing the performance of each functions involved in \ac{NR} \ac{PF}, these can be categorized as following:  
\begin{itemize}
  \item FP Dominant (Float): Calculate \ac{NPM}, Update $d\{P,Q\}/d\{\angle V,|V|\}$
  \item Memory and FP (LU Total): LU Refactorization, FS-BS
  \item Memory Dominant (Memory): Permute \ac{JM}($A$), Update $V$
\end{itemize}

The performance on CPU is evaluated with timing and Intel Vtune profiling tool for the memory access pattern analysis. \cref{fig:kernel_saturation_cpu} shows the time of each process of SMT parallelization and SMT+SIMD parallelization with different number of threads. For both small and mid-size grid, similar to the observation in batched linear solver, the performance increased almost linear at the beginning and only slightly after 6 threads. For the mid-size grid, with the last-level cache missing rate of the SMT+SIMD with 12 threads increased from 0.1\% in total to 5\% for "LU Total" and 25\% for "memory" comparing to with only 4 threads. The random memory access pattern of "LU Total" and "memory" requires a frequent exchange between the cache and system DRAM, which due to its limited bandwidth actually, drags the overall performance back by increasing thread number. The SMT+SIMD version still outperforms 12 threaded SMT parallelization scheme at the thread number of 5. For the small grid, with the data fits in the cache well, the increase of threads number improves the overall performance.

Concluding, on CPU platform, increasing number of threads brings only benefit, when solving \ac{PF} of a small scale grid. While with SIMD+SMT, the cache should be considered and the number of threads should be carefully chosen given the grid size and available hardware. On the test CPU for instance, the number of the physical cores (in this case 6) is a good first trial.

\subsection{Performance analysis on GPU}

\begin{table*}[hbt!]
  \caption{Benchmark results profile simulation including initialization}
  \centering
  \begin{tabular}{@{}lrrrrrrrr@{}}
    \toprule
     case name & case118  & case300  & case1354 & case2869 & case9241  & sb mv-lv & sb hv-mv & sb ehv-hv \\ 
     number of buses & 118 & 300 & 1354 & 2869 & 9241 & 115 & 1787 & 713 \\
     number of branches & 186& 411 & 1991& 4582 & 16019 & 115 & 1836 & 1275 \\
     \midrule
     \multicolumn{9}{c}{Timing of the initialization (ms)} \\
        & 24     & 67     & 267    & 608    & 2211    & 36     & 353    & 142  \\
    \midrule
    \multicolumn{9}{c}{Timing of 10,000 \ac{PF}s (ms)} \\

    pandapower & 18,956 &	47,373	 & 149,234 &	464,009 &	1,794,955 &	12,366	& 126,062 &	101,279      \\    
    SMT best  & 175 & 778    & 2,666   & 8,323   & 40,835   & 107    & 2,291   & 1,558    \\
    SMT+SIMD best & 97 & 333   & 1,524   & 7,217   & 38,848   & 52    & 1,646   & 721    \\
    1 gpu best & 50  & 201    & 639    & 1,941   & 10,658   & 43    & 656   & 449    \\
    2 gpu best & 26 & 102    & 342    & 1,021   & 6,525    & 24     & 353   & 233    \\ \bottomrule
    \end{tabular}

  \label{tab:final_benchmarking}
\end{table*}

The GPU version requires the extra memory transaction between host and device, which is labelled as "DH Memory". \cref{fig:kernel_saturation_gpu} shows the function performance on GPU with one stream for both grids with 10,000 calculations. It can be observed that except for the "LU Total", the other tasks can saturate the GPU resources with small batch size, thus it is insensitive to the change of the batch size for both grids. Comparing to the the best CPU performance, the "\ac{FP}" function achieved the average improvement among x5-6.
\begin{figure}[hbt!]
  \centering
  \includegraphics[width=0.48\textwidth]{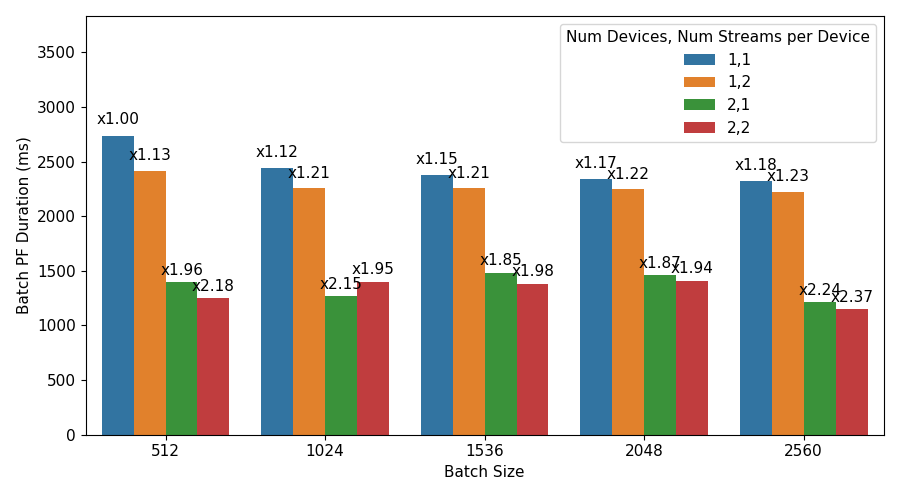}
  \caption{Duration and speedups with batches and number of streams with case2869pegase.}
  \label{fig:batchsize_numdev_numstream}
\end{figure}

\cref{fig:batchsize_numdev_numstream} shows the improvement with the usage of CUDA concurrency with stream and multi GPUs on different batch sizes with case2869pegase. The test case with batch size 512 and 1 stream is used as base line scenario. With the multiple streams on one GPU, an improvement due to the hiding the memory transaction between host and device can be observed with all batch sizes. When the batch size is small, the improvement is higher which is related to the potential of kernel overlapping. When two GPUs are available, an acceleration of around factor x2 can be observed. However, due to the imbalanced computation load distribution among GPUs and streams. With specific stream/GPU undertakes more tasks than the others, the improvement of the overall execution time varies. Using batch size 2048 as an example, when executing 10,000 \ac{PF}s, 4x2048 tasks are equally distributed, while one GPU/Stream have to execute the 1808 leftovers.

\subsection{Benchmarking result of the integrated parallel PF-solver} 

\begin{figure}[hbt!]
  \centering
  \includegraphics[width=0.48\textwidth]{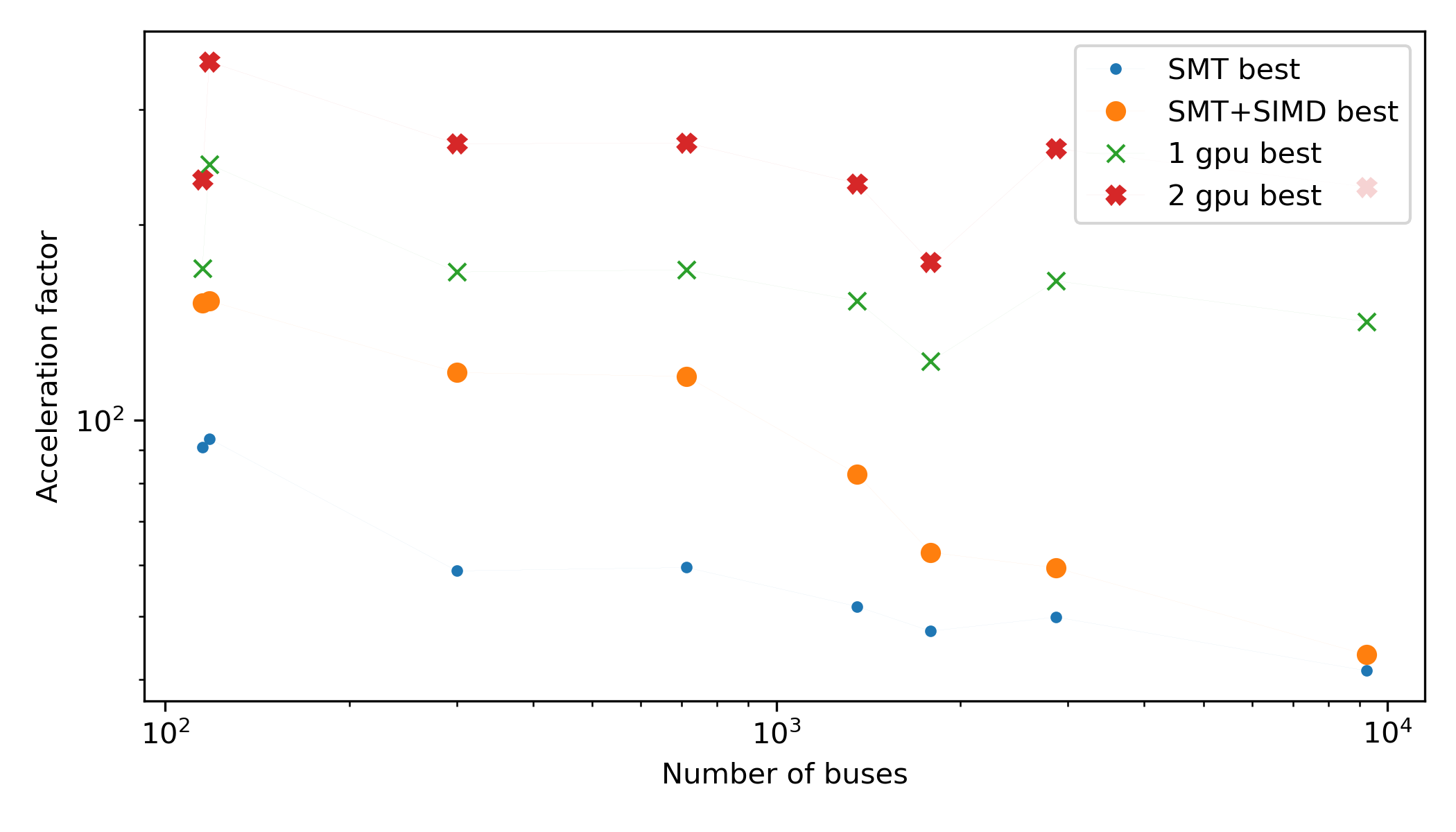}
  \caption{Acceleration factor of proposed approach comparing to baseline case pandapower including initialization.}
  \label{fig:acce_fact}
\end{figure}

The benchmarking result of running 10,000 \ac{PF}s of multiple grids with same $Y_{bus}$ is listed in \cref{tab:final_benchmarking}. The parallel \ac{PF}-solver is integrated in pandapower. Besides case2869pegase and IEEE case300, further standard grids "IEEE case118", "case1354pegase" and "case9241pegase" available in pandapower are utilized to evaluate the performance under different grid dimensions. As well as three grids from the SimBench open-source dataset\cite{Meinecke.2020}, which was recreated to represent the characteristics of real German grid and contains realistic time series data for loads and \ac{DER}s, are used for benchmark. The dataset contains yearly time series of 15-minutes resolution (35040 total time steps), the first 10,000 time steps are used. With multiple performance optimizations especially the creation of \ac{JM}, pandapower gives already better performance comparing to MATPOWER \cite{Thurner.2018,Schafer.}, thus the baseline case is performed with \ac{PF}-function of pandapower named "newtonpf", which is not parallelized and used SuperLU as linear system solver.

 \cref{fig:acce_fact} shows the acceleration factor with regard to the number of buses in the grid. The acceleration factor is calculated by $T_{pandapower} / T_{case}$. It would be fair for the evaluation of the gained acceleration with the comparison against baseline, since the required \ac{NR} iterations and the grid topology are the same in pandapower and in the proposed approach. As can also be seen in \cref{tab:final_benchmarking}, with the increase of grid dimension, the acceleration on the GPU is more significant as on CPU, due to the high \ac{FP} capability and memory bandwidth of GPU. On large grids, e.g. case9241pegase, the GPU version is x4 times faster than the best CPU performance. To be noticed by case9241pegase, due to the large requirements of GPU memory, batch size larger than 2048 failed in the benchmarking. The effect of acceleration by using SMT+SIMD is also decreasing comparing to \ac{SMT} due to the aforementioned cache issue. On small size grid, because of the well usage of CPU cache, the performance of CPU is satisfying. 

 From \cref{tab:final_benchmarking}, it can be further observed, that the grid dimension has direct impact on the one-time initialization time, which is according to \cref{fig_process_overview} applied on both CPU and GPU cases. The one-time initialization takes significant amount of time, since most of the code is executed in Python environment, which can be further optimized e.g. with \ac{JIT} compilation or C/C++ integration.

In the test considering the number of \ac{PF} calculation, when increasing the number of calculation from 100 to 10,000, the number of calculation has few impact on the average time of solving single \ac{PF} on CPU SMT and SMT+SIMD. The GPU version, due to the aforementioned saturation issue, the performance is improved with the increase of the number of calculation. After the saturation point of around 2000 calculations is reached, the average time of solving \ac{PF} is almost constant.

%% file: conclusion.tex
In this paper, an approach for the parallel solving of many power flows with CPU/GPU is proposed. When the grid admittance matrices share the same sparsity pattern, the Newton-Raphson power flows can be efficiently solved in parallel. The performance is evaluated in detail with multiple test cases covering small, mid-size and large grids.

Impressive acceleration (more than 100x over single threaded open-source tool pandapower and at least 16x can be expected if pandapower can be perfectly parallelized on 6 cores) was achieved with CPU/GPU parallelization. The performance of the fast parallel power flow solver originated mainly from the following points:
\begin{itemize}
    \item Avoidance of repetitive work (sparse matrix indexing initialization, pre-ordering, lookup creation, etc.),
    \item Reduction of computational overhead with LU-refactorization,
    \item Hardware-specific optimization and parallelization strategies.
\end{itemize}
In detail, on CPU platform, the power flows can be accelerated with less effort with SMT parallelization. With SMT+SIMD parallelization, the acceleration effect is highly dependent to the problem scaling. On small grids, a further speed-up of x2-3 comparing to SMT parallelization can be expected. On the GPU platform, with the batch operation on sparse matrix, the high \ac{FP} capability and high memory bandwidth can be effectively saturated. Comparing to the CPU counterparts, the computing-bounding functions are accelerated significantly, while the memory-bounding functions depend highly on the problem scaling. 

The outstanding performance of the proposed parallel power flow solver shows promising application in the real-time grid operation in order to allow the consideration of uncertainties. The innovative researches in the data-driven machine-learning methods in power systems can be great benefitted. Even more potential can be exploited with the application of the proposed solver on a high-performance computing clusters with multiple CPUs and GPUs.